\documentclass[12pt]{article}
\usepackage{graphicx}
\usepackage{amssymb, subfig}
\usepackage{color}
\usepackage{float}
\usepackage{amsmath,verbatim,amssymb,epsfig,indentfirst,graphicx,multirow,color,graphics,rotating,url}
\usepackage{bm}
\usepackage{natbib,theorem}
\newcommand{\red}{\textcolor{black}}
\newcommand{\blue}{\textcolor{black}}

\def\v{{\varepsilon}}
\def\bv{{ \boldsymbol{\varepsilon}}}

\newtheorem{theorem}{Theorem}
\newtheorem{lemma}{Lemma}

\newtheorem{proposition}{Proposition}

\newtheorem{example}{Example}
\newcommand{\bA}{\mathbf{A}}

\newcommand{\bB}{\mathbf{B}}

\newcommand{\bE}{\mathbf{E}}
\newcommand{\be}{\mathbf{e}}

\newcommand{\bI}{\mathbf{I}}

\newcommand{\bP}{\mathbf{P}}

\newcommand{\bU}{\mathbf{U}}

\newcommand{\bbv}{\mathbf{v}}

\newcommand{\bX}{\mathbf{X}}
\newcommand{\bx}{\mathbf{x}}

\newcommand{\by}{\mathbf{y}}

\newcommand{\bz}{\mathbf{z}}

\newcommand{\cN}{\mathcal{N}}

\newcommand{\cM}{\mathcal{M}}

\newcommand{\cS}{\mathcal{S}}
\newcommand{\cT}{\mathcal{T}}

\newcommand{\cI}{\mathcal{I}}

\newcommand{\eps}{\varepsilon}

\newcommand{\bbeta}{\boldsymbol{\beta}}

\newcommand{\bomega}{\boldsymbol{\omega}}

\newcommand{\bSigma}{\boldsymbol{\Sigma}}

\newcommand{\btheta}{\boldsymbol{\theta}}

\newcommand{\E}{\mathrm{E}}
\newcommand{\Var}{\mathrm{Var}}

\newcommand{\tr}{\mathrm{tr}}
\newcommand{\sign}{\mathrm{sign}}
\newcommand{\bzero}{\boldsymbol{0}}
\newcommand{\bone}{\boldsymbol{1}}

\newcommand{\argmin}{\operatornamewithlimits{argmin}}

\def\argmin{\mathop{\rm argmin}}

\newcommand{\bea}{\begin{eqnarray*}}
\newcommand{\eea}{\end{eqnarray*}}
\newcommand{\bay}{\begin{array}}
\newcommand{\eay}{\end{array}}
\newcommand{\ben}{\begin{enumerate}}
\newcommand{\een}{\end{enumerate}}
\newcommand{\bcen}{\begin{center}}
\newcommand{\ecen}{\end{center}}

\newcommand{\bb}{\mathbf{b}}
\newcommand{\bu}{\mathbf{u}}

\newcommand{\cA}{\mathcal{A}}
\newcommand{\cB}{\mathcal{B}}

\newcommand{\cH}{\mathcal{H}}

\newcommand{\cSm}{\widehat\cS^{(m)}}
\newcommand{\cTm}{\widehat\cT^{(m)}}

\begin{document}
%\numberwithin{equation}{section}
%\renewcommand{\baselinestretch}{1.1}

\title{Model Selection for High Dimensional Quadratic Regression via Regularization\thanks{Ning Hao is Assistant Professor, Department of Mathematics, University of Arizona, Tucson, AZ 85721 (Email: nhao@math.arizona.edu). Yang Feng is Associate Professor, Department of Statistics, Columbia University, New York, NY 10027 (E-mail: yangfeng@stat.columbia.edu). Hao Helen Zhang is Professor, Department of Mathematics, University of Arizona, Tucson, AZ 85721 (Email: hzhang@math.arizona.edu). Ning Hao and Yang Feng contribute equally to this work.  The authors are partially supported by NSF Grants DMS-1309507 (Hao and Zhang), DMS-1308566 (Feng) and DMS-1554804 (Feng). }}%The authors are grateful to the editors, associate editor, and three referees for their helpful comments and suggestions.}}
\author{Ning Hao, Yang Feng, and Hao Helen Zhang}
%\\University of Arizona and Columbia University}

\date{}
\maketitle

%\begin{singlespace}
\begin{abstract}
Quadratic regression (QR) models naturally extend linear models by considering interaction effects between the covariates. To conduct model selection in QR, it is important to maintain the hierarchical model structure between main effects and interaction effects. Existing regularization methods generally achieve this \red{goal} by solving complex optimization problems, which usually demands high computational cost and hence are not feasible for high dimensional data. This paper focuses on scalable regularization methods for model selection in high dimensional QR. We first consider two-stage regularization methods and establish theoretical properties of the two-stage LASSO. Then, a new regularization method, called Regularization Algorithm under Marginality Principle (RAMP), is proposed to compute a hierarchy-preserving regularization solution path efficiently. Both methods are further extended to \red{solve} generalized QR models. Numerical results are also shown to demonstrate performance of the methods.
\end{abstract}


%\end{singlespace}
%\newpage
\section{Introduction}\label{S1}%%%%%%%%%%%%%%%%%%%%%%%%%%%%%%%%%%%%%%%%%%%%%%%%%%%%%

Statistical models involving two-\red{way} or higher-order interactions have been studied in various contexts, such as linear models and generalized linear models \citep{Nelder:1977,MccullaghNelder:1995}, experimental design \citep{HamadaWu:1992,ChipmanETAL:1997}, and polynomial regression \citep{Peixoto:1987}. In particular, a quadratic regression (QR) model formulated as
\begin{eqnarray}\label{AAA1}
Y=\beta_0+\beta_1X_1+\cdots+\beta_pX_p+\beta_{1,1}X_1^2+\beta_{1,2}X_1X_2+\cdots+\beta_{p,p}X_p^2+\v
\end{eqnarray}
has been considered recently to analyze high dimensional data. In (\ref{AAA1}), $X_1$,..., $X_p$ are main effects, and order-2 terms $X_jX_k$ $(1\leq j\leq k\leq p)$ include quadratic main effects ($j=k$) and two-way interaction effects ($j\neq k$). A key feature of model \eqref{AAA1} is its hierarchical structure, as order-2 terms are derived from the main effects. To reflect their relationship, we call $X_jX_k$ the \emph{child} of $X_j$ and $X_k$, and $X_j$ and $X_k$ the \emph{parents} of $X_jX_k$.

Standard techniques such as ordinary least squares can be applied to solve (\ref{AAA1}) for a small or moderate $p$. When $p$ is large and variable selection becomes necessary, it is suggested that the selected model should keep the hierarchical structure. That is, interaction terms can be selected into the model only if their parents are in the model. This is referred to the marginality principle \citep{Nelder:1977}.
%In high dimensional data analysis, variable selection is crucial for dimension reduction when the data have many features of which only a small subset are relevant to the response, see  for an overview for modern variable selection methods.
In general, a direct application of variable selection techniques to \eqref{AAA1} can not automatically ensure the hierarchical structure in the final model. Recently, several regularization methods \citep{ZhaoETAL:2009, YuanRoshanZou:2009, ChoiLiZhu:2010,bien2013lasso} have been proposed to conduct variable selection for \eqref{AAA1} under the marginality principle by designing special forms of penalty functions. These methods are feasible when $p$ is a few hundreds or less, and the resulting estimators have oracle properties when $p=o(n)$ \citep{ChoiLiZhu:2010}. However, when $p$ is much larger, these methods are not feasible since their implementation requires storing and manipulating the entire $O(p^2)\times n$ design matrix and solving complex constrained optimization problems. The memory and computational cost can be extremely high and prohibitive.

In this paper, we study regularization methods on model selection and estimation for QR and generalized quadratic regression (GQR) models under the marginality principle. The main focus is the case $p\gg n$, which is a bottleneck for the existing regularization methods. We study theoretical properties of a two-stage regularization method based on the LASSO and propose a new efficient algorithm, RAMP, which produces a hierarchy-preserving solution path. In contrast to existing regularization methods, these procedures avoid storing $O(p^2)\times n$ design matrix and sidestep complex constraints and penalties, making them feasible \red{to analyze data with many variables.} %for large $p$.
In particular, our R package \texttt{RAMP} runs well on a desktop for data with $n=400$ and $p=10^4$ and it takes less than 30 seconds  (with CPU 3.4 GHz Intel Core i7 and 32GB memory) to fit the QR model and get the whole solution path.  The main contribution of this paper is threefold. First, we establish a variable selection consistency result of the two-stage LASSO procedure for QR and offer new insights on stage-wise selection methods. %under conditions which only slightly stronger than ones required in similar results for linear regression.
To our best knowledge, this is the first selection consistency result for high dimensional QR. Second, the proposed algorithms are computationally efficient and will make a valuable contribution to interaction selection tools in practice. Third, our methods are extended to interaction selection in GQR models, which are rarely studied in literature.

We define notations used in the paper. Let $\bX=(\bx_1,...,\bx_n)^{\top}$ be the $n\times p$ design matrix of main effects and $\by=(y_1,...,y_n)^{\top}$ be the $n$-dimensional response vector. The linear term index set is $\cM=\{1,2,...,p\}$, and the order-2 index set
is $\cI=\{(j,k):1\leq j\leq k\leq p\}$. The regression coefficient vector $\bbeta=(\beta_0,\bbeta_{\cM}^{\top},\bbeta_{\cI}^{\top})^{\top}$, where $\bbeta_{\cM}=(\beta_1,...,\beta_p)^{\top}$ and $\bbeta_{\cI}=(\beta_{1,1},\beta_{1,2},...,\beta_{p,p})^{\top}$. For a subset $\cA\subset\cM$, use $\bbeta_{\cA}$ for the subvector of $\bbeta_{\cM}$ indexed in $\cA$, and $\bX_{\cA}$ for the submatrix of $\bX$ whose columns are indexed in $\cA$. \red{In particular, $\bX_j$ is the $j$th column of $\bX$.} We treat the subscripts $(j,k)$ and $(k,j)$ as identical, i.e., $\beta_{j,k}=\beta_{k,j}$. Let $c_1$, $c_2$, ... and $C_1$, $C_2$, ... be positive constants which are independent of the sample size $n$. They are locally defined and their values may vary in different context. For a vector $\bbv=(v_1,...,v_p)^{\top}$, $\|\bbv\|=\sqrt{\sum_{j=1}^pv^2_j} $ and $\|\bbv\|_1=\sum_{j=1}^p|v_j| $. For a matrix $\bA$, define $\|\bA\|_{\infty}=\max_{i}\sum_j|A_{ij}|$ and $\|\bA\|_2=\sup_{\|\bbv\|_2=1}\|\bA\bbv\|_2$ as the standard operator norm, i.e., the square root of the largest eigenvalue of $\bA^{\top}\bA$.

The rest of the paper is organized as follows. Section \ref{S2} considers two-stage regularization methods for model selection in QR and studies theoretical properties of the two-stage LASSO. Section \ref{S3} proposes the RAMP to compute the entire hierarchy-preserving solution path efficiently. Section \ref{S4} discusses the generalizations of the proposed methods to GQR models. Section \ref{S5} presents numerical studies, followed by a discussion. Technical proofs are in the Appendix.

\section{Two-stage Regularization Method} \label{S2} %%%%%%%%%%%%%%%%%%%%%%%%%%%%%%%%%%%%%
Variable selection and estimation via penalization is popular in high dimensional analysis. Examples include the LASSO \citep{TibshiraniLASSO:1996}, SCAD \citep{FanLiSCAD:2001}, elastic net \citep{ZouHastie:ENET:2005}, minimax concave penalty (MCP) \citep{ZhangMCP:2010}, among many others. Properties such as model selection consistency and oracle properties have been verified
%for these methods even if $p$ grows exponentially fast as $n$ grows
\citep{Zhao:Yu:2006,wainwright2009sharp,FanLvSCAD:2011}.
A general penalized estimator for linear models is defined as
\begin{eqnarray}\label{AAA2}
(\hat\beta_0,\hat\bbeta_{\cM})=\argmin_{(\beta_0,\bbeta_{\cM})} \frac{1}{2n}\|\by-\bone\beta_0-\bX\bbeta_{\cM}\|^2+\sum_{j=1}^p J_{\lambda}(\beta_j),
\end{eqnarray}
where $\by$ is the response vector, $\bX$ is the design matrix, $J_{\lambda}(\cdot)$ is a penalty function, and $\lambda\geq 0$ is a regularization parameter. The penalty $J(\cdot)$ and $\lambda$ may depend on index $j$. For easy presentation, we use same penalty function and parameter for all $j$ unless stated otherwise.

We consider the problem of variable selection for QR model (\ref{AAA1}). Define $\bX^{\circ2}=\bX\circ\bX$ as an $n\times\frac{p(p+1)}{2}$
matrix consisting of all pairwise column products. That is, for $\bX=(\bX_1,...,\bX_p)$, \red{$\bX^{\circ2}=\bX\circ\bX=(\bX_1\star\bX_1,\bX_1\star\bX_2,...,\bX_p\star\bX_p)$, where $\star$ denotes} the entry-wise product of \blue{two} column vectors. \red{For an index set $\cA\subset\cM$, d}efine $\cA^{\circ2}=\cA\circ\cA=\{(j,k): j\leq k; j,k\in\cA\}\subset\cI$\red{, and $\cA\circ\cM=\{(j,k): j\leq k; j \text{ or }k\in\cA\}\subset\cI$.} We use $\bX_{\cA}^{\circ2}$ as a short notation for $(\bX_{\cA})^{\circ2}$, a matrix whose columns are indexed by $\cA^{\circ2}$.
%\red{The set of important linear terms is denoted by $\cS=\{j: \beta_j\ne 0, j=1, \ldots, p\}$}.

Two-stage regularization methods for interaction selection have been considered in \cite{LARS:2004,WuTETAL:2009}, among others. However, their theoretical properties are not clearly understood. In the following, we first illustrate the general two-stage procedure for interaction selection.

\vspace{3mm}
\noindent
{\it Two-stage Regularization Method}:

Stage 1: Solve (\ref{AAA2}). Denote the selected model by $\widehat{\cA}=\{j:\hat\beta_j\neq0, j=1,\ldots, p\}$.

Stage 2: Solve
\begin{eqnarray*}
\hat\bbeta=\argmin_{\bbeta} \frac{1}{2n} \|\by-\bone\beta_0-\bX_{\widehat{\cA}}\bbeta_{\widehat{\cA}}-\bX_{\widehat{\cA}}^{\circ2}\bbeta_{\widehat{\cA}^{\circ2}}\|^2%+\sum_{j\in\widehat{\cA}} J_{\lambda}(\beta_j)
+\sum_{\alpha\in\widehat{\cA}^{\circ2}} J_{\lambda}(\beta_{\alpha}),
\end{eqnarray*}

At Stage 1, only main effects are considered for selection, with all the order-2 terms being left out of the model. Denote the selected set by $\widehat{\cA}$. At Stage 2, we expand $\widehat\cA$ by including all the two-way interactions of those main effects within $\widehat\cA$ and fit the new model. To keep the hierarchical structure, we do not penalize main effects at Stage 2, i.e., set $J_{\lambda}(\cdot)=0$ for $j\in\widehat\cA$. In order to keep the hierarchy, it is also possible to use other methods \citep{ZhaoETAL:2009, YuanRoshanZou:2009, ChoiLiZhu:2010,bien2013lasso} at Stage 2.

One main advantage of this two-stage regularization procedure is its simple implementation. Existing R packages \texttt{lars} and \texttt{glmnet} can be directly used to carry out the procedure. %This is contrast to existing methods which use special penalty functions and demand high computational cost.
Stage 1 \red{serves as} %is actually
a dimension reduction step prior to Stage 2, so the two-stage method avoids estimating $O(p^2)$ parameters altogether, making the procedure feasible for very large $p$.

%\subsection{Model Selection Consistency of the Two-Stage LASSO}
In spite of its computational advantages, theoretical properties of two-stage regularization methods are seldom studied in literature. A commonly raised concern is whether the important main effects can be consistently identified at Stage 1, when all order-2 terms are left out of the model on purpose. Next, we focus on the two-stage LASSO method and investigate its selection behavior at Stage 1. In particular, we establish the main-effect selection consistency result of the two-stage LASSO for QR under some regularity conditions.
%which \red{are} only slightly stronger than \red{simple linear regression models.}
%Although it is intuitive from former discussion, we formally establish the main effect selection consistency in Stage 1 for a two-stage LASSO method. By no means we put the technical conditions weakest possible. However, from the proof, it is easy to see that similar conclusions hold under weaker conditions and different penalties, and for more flexible models such as GLMs. It is only technically more difficult to generalize the result to the more general settings, say, the framework introduced in \cite{FanLvSCAD:2011}.

The LASSO is a special case of (\ref{AAA2}) by using the $\ell_1$ penalty
%$J_{\lambda}(\beta_j)=\lambda|\beta_j|$.
\begin{eqnarray*}
(\hat\beta_{0L},\hat\bbeta_L)=\argmin_{(\beta_0,\bbeta_{\cM})} \frac{1}{2n}\|\by-\bone\beta_0-\bX\bbeta_{\cM}\|^2+\sum_{j=1}^p \lambda|\beta_j|.
\end{eqnarray*}
In the following, we show
%the main effect selection consistency at Stage 1 for the two-stage LASSO approach. In particular, we show
that the LASSO solution $\hat\bbeta_L$ is sign consistent at Stage 1, i.e., $\sign(\hat\bbeta_L)=\sign(\bbeta_{\cM})$ with an overwhelming probability for a properly chosen tuning parameter. This result provides critical theoretical insight about the two-stage LASSO estimator.

Consider a sparse quadratic model with a Gaussian design. % which generalizes the set up in \cite{wainwright2009sharp}.
Assume that $\bx_i$, $1\leq i\leq n$, are independent and identically distributed (i.i.d.) from $\cN(\bzero,\Sigma)$, and
\begin{eqnarray}\label{AAA3}
y_i=\beta_0+\bx_i^{\top}\bbeta_{\cM}+(\bx_i^{\top})^{\circ2}\bbeta_{\cI}+\eps_i,
\end{eqnarray}
where $\bv=(\eps_1,...,\eps_n)^{\top} \sim \cN(\bzero,\sigma^2\bI)$ is independent of $\{\bx_i\}_{i=1}^n$.   Without loss of generality, we further center $y_i$ and $(\bx_i^{\top})^{\circ2}$ and write
\begin{eqnarray}\label{AAA4}
y_i=\bx_i^{\top}\bbeta_{\cM}+\bu_i^{\top}\bbeta_{\cI}+\eps_i,
\end{eqnarray}
where $y_i$ is the centered response and $\bu_i^{\top}=(\bx_i^{\top})^{\circ2}-\E(\bx_i^{\top})^{\circ2}$ is a $p\times (p+1)/2$ \red{row} vector with all centered order-2 terms. %We require that main effects are centered in (\ref{AAA3}) to make $\sign(\bbeta_{\cM})$ well-defined; see \cite{HaoZhang:note:2012} for further explanations on the well-definedness of sign and support of the coefficient vector $\bbeta_{\cM}$ for a QR model.
Let $y_{\cM i}=\bx_i^{\top}\bbeta_{\cM}$ and $y_{\cI i}=\bu_i^{\top}\bbeta_{\cI}$. $\by_{\cM}=(y_{\cM 1,},...,y_{\cM n})^{\top}$, $\by_{\cI}=(y_{\cI 1,},...,y_{\cI n})^{\top}$, $\bU=(\bu_1,...,\bu_n)^{\top}$. Set %$\eps_i\sim \cN(0,\sigma^2)$, and
$\tau^2=\Var(y_{\cI i})$. Define $\omega_i=\bu_i^{\top}\bbeta_{\cI}+\eps_i$ and $\bomega=(\omega_1,...,\omega_n)^{\top}$\red{, which is treated as noise at Stage 1.}  Denote by $\Sigma_{\cA\cB}$ the submatrix of $\Sigma$ with row index $\cA$ and column index $\cB$. \red{As illustrated in \cite{HaoZhang:note:2012}, the support and sign of the coefficient vector $\bbeta_{\cM}$ for a QR model depend on} \blue{its} \red{parametrization because a coding transformation can change the support of $\bbeta_{\cM}$. Therefore, we follow \cite{HaoZhang:note:2012} and define the index set of important main effects by $\cS=\{j: \beta_j^2+\sum_{k=1}^p\beta_{j,k}^2> 0\}$. Let $s=|\cS|$ and $\cT=\{(k,\ell):\beta_{k,\ell}\ne0\}.$ It follows this definition that $\cT\subset\cS^{\circ2}$. Moreover, in order to make $\sign(\bbeta_{\cM})$ well-defined, we require that main effects are centered in (\ref{AAA3}). We refer to \cite{HaoZhang:note:2012} for further explanations on the well-definedness of sign and support of the coefficient vector $\bbeta_{\cM}$ for a QR model.}

Define $\Sigma_{\cS^c\mid\cS}=\Sigma_{\cS^c\cS^c}-\Sigma_{\cS^c\cS}(\Sigma_{\cS \cS })^{-1}\Sigma_{\cS \cS^c}$ where $\cS^c=\cM-\cS$. Let $\Lambda_{\min}(\bA)$ be the smallest eigenvalue of $\bA$ and $\rho_u(A)=\max_iA_{ii}$. Assume the following technical conditions:
 \begin{description}
   \item[(C1)] (Irrepresentable Condition) $\|\Sigma_{\cS^c\cS}(\Sigma_{\cS\cS})^{-1}\|_{\infty}\leq 1-\gamma,$ $\gamma\in (0,1]$.
   \item[(C2)] (Eigenvalue Condition) $\Lambda_{\min}(\Sigma_{\cS\cS})\geq C_{\min}>0$.
  % \item[(C3)] (Heredity Condition) If $\beta_{jk}\ne0$, then $\beta_j\beta_k\ne 0$.
   %\item[C4] (Dimensionality and Beta-min Condition) $\log p=O(n^{\xi_1})$, $s=O(n^{\xi_2})$, $\rho_{u}(\Sigma_{\cS^c|\cS})=o(\xi_3)$, with positive constants $\xi_1$, $\xi_2$, $\xi_3$, and $\xi_1+2\xi_2+\xi_3<1$; $\beta_{\min}=\min_{j\in\cS}|\beta_j|>$.
 \end{description}

\begin{theorem} \label{th::two.stage.lasso}
Consider the quadratic model with a random Gaussian design (\ref{AAA4}). Suppose that (C1)-(C2) hold. Consider the family of regularization parameters
\begin{eqnarray}\label{AAA5}
\lambda_n(\phi_p)=\sqrt{\frac{\phi_p\rho_u(\Sigma_{\cS^c|\cS})}{\gamma^2}\frac{4(\sigma^2+\tau^2)\log(p)}{n}}
\end{eqnarray}
for some $\phi_p\geq 2$. If for some fixed $\delta>0$, the sequence $(n,p,s)$ and regularization sequence $\{\lambda_n\}$ satisfy
\begin{eqnarray}\label{AAA6}
\frac{n}{2s\log(p-s)}>(1+\delta)\frac{\rho_u(\Sigma_{\cS^c|\cS})}{C_{\min}\gamma^2}(1+\frac{2(\sigma^2+\tau^2)C_{\min}}{\lambda^2_ns}),
\end{eqnarray}
then the following holds with probability greater than $1-c_1\exp(-c_2\min\{s,\log(p-s),n^{\frac12}\})$.
\begin{enumerate}
\item The LASSO has a unique solution $\hat\bbeta_L$ with support contained within $\cS$.
\item Define the gap
\begin{eqnarray}\label{AAA7}
g(\lambda_n)=c_3\lambda_n\left\|\Sigma_{\cS\cS}^{-\frac12}\right\|^2_{\infty}+20\sqrt{\frac{\sigma^2 s}{C_{\min}n}}+\frac{9\|\bbeta_{\cI}\|_2\sqrt{s}}{C_{\min}n^{\frac13}}.
\end{eqnarray}
Then if $\beta_{\min}=\min_{j\in\cS}|\beta_j|>g(\lambda_n)$, then $\sign(\hat\bbeta_L)=\sign(\bbeta_{\cM})$.
\end{enumerate}
Furthermore, given (\ref{AAA5}), an alternative condition to (\ref{AAA6}) making the above results hold is
\begin{eqnarray}\label{AAA8}
\frac{n}{2s\log(p-s)}>\frac{1+\delta'}{1-\frac{1}{\phi_p}}\frac{\rho_u(\Sigma_{\cS^c|\cS})}{C_{\min}\gamma^2}
\end{eqnarray}
for some $\delta'>0$.
\end{theorem}

\medskip

Remark 1. Conditions (C1)-(C2) are commonly used to show model selection consistency of the LASSO estimator in the literature. Conditions (\ref{AAA6}) and (\ref{AAA7}) are key requirements on dimensionality and minimal signal strength $\beta_{\min}$, respectively. \red{The normality assumption is used here to facilitate \blue{proof} and comparison to existing results \blue{in} linear regression. In the supplementary material, we \blue{establish} Theorem 2 which extends the consistency result to \red{the} non-Gaussian designs. \blue{Other} possible extensions \blue{of theoretical results} are discussed in Section \ref{S6}.}

%A relaxed version of Theorem 1 is illustrated below.%In particular, \blue{they} are similar to those used to prove the sign consistency of the lasso for linear models with a Gaussian random design, e.g., Theorem 3 of \cite{wainwright2009sharp}. Here slightly stronger conditions are imposed on dimensionality and $\beta_{\min}$ than in standard linear regression cases, as the noise $\omega_i$ is not Gaussian and independent of $\bx_i$ at Stage 1.
%\begin{corollary}
%\end{corollary}

\medskip

Remark 2. The result in Theorem 1 generalizes Theorem 3 in \cite{wainwright2009sharp} that is established in the context of linear regression. Theorem 1 implies that the two-stage LASSO can identify \red{important} main effects at Stage 1. The validity of the two-stage LASSO is then guaranteed \red{as the index set of important interactions $\cT\subset\cS^{\circ2}$. That is,} all important interaction effects can be included at Stage 2. \blue{Given the result of Theorem 1}, the interaction selection consistency result of Stage 2 can be obtained under some mild conditions on the matrix $\bX^{\circ2}_{\cS}$, since the data dimensionality has been greatly reduced. \red{One can also apply existing methods, e.g., \cite{ChoiLiZhu:2010} at Stage 2, for which the selection consistency has been established.} %by the strong heredity condition \citep{YuanRoshanZou:2009, ChoiLiZhu:2010}. That is, if $\beta_{j,k}\ne0$, then $\beta_j\beta_k\ne 0$. It implies that parents of important interaction effects are nontrivial, and based on Theorem 1, they can be selected with high probability at Stage 1. Consequently, all important interaction effects can be included at Stage 2. After Theorem 1 is established, the interaction selection consistency result of Stage 2 can be obtained under some mild conditions on the matrix $\bX^{\circ2}_{\cS}$, since the data dimensionality has been greatly reduced.

%original remark 2.
%Remark 2. The result in Theorem 1 generalizes Theorem 3 in \cite{wainwright2009sharp} that is established in the context of linear regression. Theorem 1 essentially implies that the two-stage LASSO can identify nontrivial main effects at stage one. The validity of the two-stage LASSO is then guaranteed by the strong heredity condition \citep{YuanRoshanZou:2009, ChoiLiZhu:2010}. That is, if $\beta_{j,k}\ne0$, then $\beta_j\beta_k\ne 0$. It implies that parents of important interaction effects are nontrivial, and based on Theorem 1, they can be selected with high probability at Stage 1. Consequently, all important interaction effects can be included at Stage 2. After Theorem 1 is established, the interaction selection consistency result of Stage 2 can be obtained under some mild conditions on the matrix $\bX^{\circ2}_{\cS}$, since the data dimensionality has been greatly reduced.

%\red{***Ning, add a reference for finite dimensional model selection***}
%See \cite{HaoZhang:note:2012} for detail. %Remark. In particular, they are similar to those used to prove the sign consistency of the lasso for linear models with a Gaussian random design, e.g., Theorem 3 of \cite{wainwright2009sharp}. Here slightly stronger conditions are imposed on dimensionality and $\beta_{\min}$ than in standard linear regression cases, as the noise $\omega_i$ is not Gaussian and independent of $\bx_i$ at Stage 1.}

\section{Regularization Path Algorithm under Marginality Principle (RAMP)}\label{S3}%Build-in LASSO. (SCAD)

For linear regression models, regularization solution-path algorithms provide state-of-the-art computational tools to implement variable selection with high dimensional data. Popular algorithms include least angle regression (LARS) \citep{LARS:2004}, its extensions \citep{park2007l1, wu2011ordinary,zhou2012generic}, and coordinate decent algorithm (CDA) \citep{CDA:2007,WuLange:2008,friedman2010regularization,Yu.Feng.2013}. \red{These} \blue{computational} \red{tools can be used to implement two-stage methods} \blue{for fitting} QR. \red{However, by the nature of two-stage approach, the whole solution-path highly depends on the selection result at Stage 1, which is obtained under considerably high noise level} \blue{if} \red{ interaction effects are strong. Therefore, it is desirable to develop a seamless path algorithm which can select main and interaction effects simultaneously while keeping the hierarchy structure.}
%To select interactions in QR under the marginality principle, a hierarchy-preserving solution path algorithm would be useful. The problem is challenging for $p\gg n$.
\red{To achieve the goal, we propose} a Regularization Algorithm under Marginality Principle (RAMP) via \blue{the} coordinate descent to compute the solution path \blue{while} preserving the model hierarchy along the path.

We first review the coordinate decent algorithm for the standard LASSO. Consider
\begin{eqnarray*}
\min  \frac{1}{2n}\sum_{i=1}^n(y_i-\beta_0-\bx_i^{\top}\bbeta_{\cM})^2+\lambda\|\bbeta_{\cM}\|_1.
\end{eqnarray*}
There exists a penalty parameter $\lambda_{\max}$ such that the minimizer $\hat\bbeta_L=\bzero$ if $\lambda\geq\lambda_{\max}$. As $\lambda$ decreases from $\lambda_{\max}$ to 0, the LASSO solution $\hat\bbeta_L=\hat\bbeta_{\lambda}$ changes from $\bzero$ to the least squares estimator (if it exists). Usually, a sequence of values $\{\lambda_k\}_{k=1}^K$ between $\lambda_{\max}$ and $\zeta\lambda_{\max}$ is set, with $0<\zeta<1$, and a solution path $\hat\bbeta_{\lambda_k}$ is calculated for each $\lambda_k$. For a fixed $k$, using $\hat\bbeta_{\lambda_{k-1}}$ as the initial value, the CDA solves the optimization problem by cyclically minimizing each coordinate $\beta_j$ until convergence. Define $\cM_{k}= supp\{\hat\bbeta_{\lambda_k}\}$, i.e., the active set for each $\lambda_k$.

In the following, we propose a coordinate descent algorithm to fit the quadratic model under regularization which obeys the marginality principle. Given a tuning parameter $\lambda$, the algorithm computes the $\ell_1$ regression coefficients of main effects and interactions subject to the heredity condition. At step $k-1$, denote the current active main effect set as $\cM_{k-1}$ and the interaction effect set as $\cI_{k-1}$. Define $\cH_{k-1}$ as the parent set of $\cI_{k-1}$, i.e., it contains the main effect which has at least one interaction effect (child) in $\cI_{k-1}$. Set $\cH_{k-1}^c=\cM-\cH_{k-1}$.

\bigskip
\noindent
{\bf Regularization Algorithm under Marginality Principle (RAMP)}:

\vspace{2mm}
{\it Initialization}: Set $\lambda_{\max}=n^{-1}\max |\bX^{\top}\by|$ and $\lambda_{\min}=\zeta\lambda_{\max}$ with some small $\zeta>0$.  Generate an exponentially decaying sequence $\lambda_{\max}=\lambda_1>\lambda_2>\cdots>\lambda_K=\lambda_{\min}$.  Initialize the main effect set $\cM_{0}=\emptyset $ and the interaction effect set $\cI_{0}=\emptyset$.

\vspace{2mm}
{\it Path-building}: Repeat the following steps for $k=1,\cdots,K$. Given $\cM_{k-1}, \cI_{k-1}, \cH_{k-1}$, add the possible interactions among main effects in $\cM_{k-1}$ to the current model. Then with respect to $(\beta_0,\bbeta_{\cM}^{\top},\bbeta_{\cM_{k-1}^{\circ2}}^{\top})^{\top}$, we minimize
%\begin{eqnarray}
%\frac{1}{2n}\sum_{i=1}^n\left(y_i-\beta_0-\bx_i^{\top}\bbeta_{\cM}-(\bx_i^{\top})_{\cM_{{k-1}}}^{\circ2}\bbeta_{\cM_{{k-1}}^{\circ2}}\right)^2+
%\lambda_k\left[\sum_{j\in \cH^c_{{k-1}}} |\beta_j|+ \|\bbeta_{\cM_{k-1}^{\circ2}}\|_1\right],
%\end{eqnarray}
\begin{eqnarray}\label{new9}
\frac{1}{2n}\sum_{i=1}^n\left(y_i-\beta_0-\bx_i^{\top}\bbeta_{\cM}-(\bx_i^{\top})_{\cM_{{k-1}}}^{\circ2}\bbeta_{\cM_{{k-1}}^{\circ2}}\right)^2+
\lambda_k  \|\bbeta_{\cH^c_{{k-1}}}\|_1+ \lambda_k\|\bbeta_{\cM_{k-1}^{\circ2}}\|_1 ,
\end{eqnarray}
where the penalty is imposed on the candidate interaction effects and $\cH^c_{k-1}$, which contains the main effects not enforced by the strong heredity constraint. Record $\cM_k$, $\cI_k$ and $\cH_k$  according to the solution. Add the corresponding main effects from $\cI_k$ into $\cM_k$ to enforce the heredity constraint, and calculate the OLS based on the current model. %{\bf Ning, should we say one sentence about OLS here?}

\red{Compared with the two-stage approach}, \blue{the} \red{RAMP allows at each step the interaction effects $\cM_{k-1}^{\circ2}$ to enter the model for selection. Following the same strategy, we propose a weak hierarchy  version of RAMP, denoted by RAMP-w, as a flexible relaxation. The main difference is that we use the set $\cM_{k-1}\circ\cM$ instead of $\cM_{k-1}^{\circ2}$ in (\ref{new9}) and solve the optimization problem with respect to $(\beta_0,\bbeta_{\cM}^{\top},\bbeta_{\cM_{k-1}\circ\cM}^{\top})^{\top}$. It is clear that an interaction effect can enter the model for selection} \blue{as long as} \red{one of its parents is selected in a previous step. Therefore, it is helpful in the scenario when only one of the parents of an important effect is strong. Both versions of RAMP} \blue{are} \red{implemented in our R package \texttt{RAMP}}, \blue{which is} \red{available on} \blue{the} \red{CRAN web site for researchers to use. Moreover, other penalty options such as SCAD and MCP are also included} \blue{in the \texttt{RAMP} package}.

\begin{figure}[H] %figure 1
\begin{center}
\includegraphics[scale=0.4]{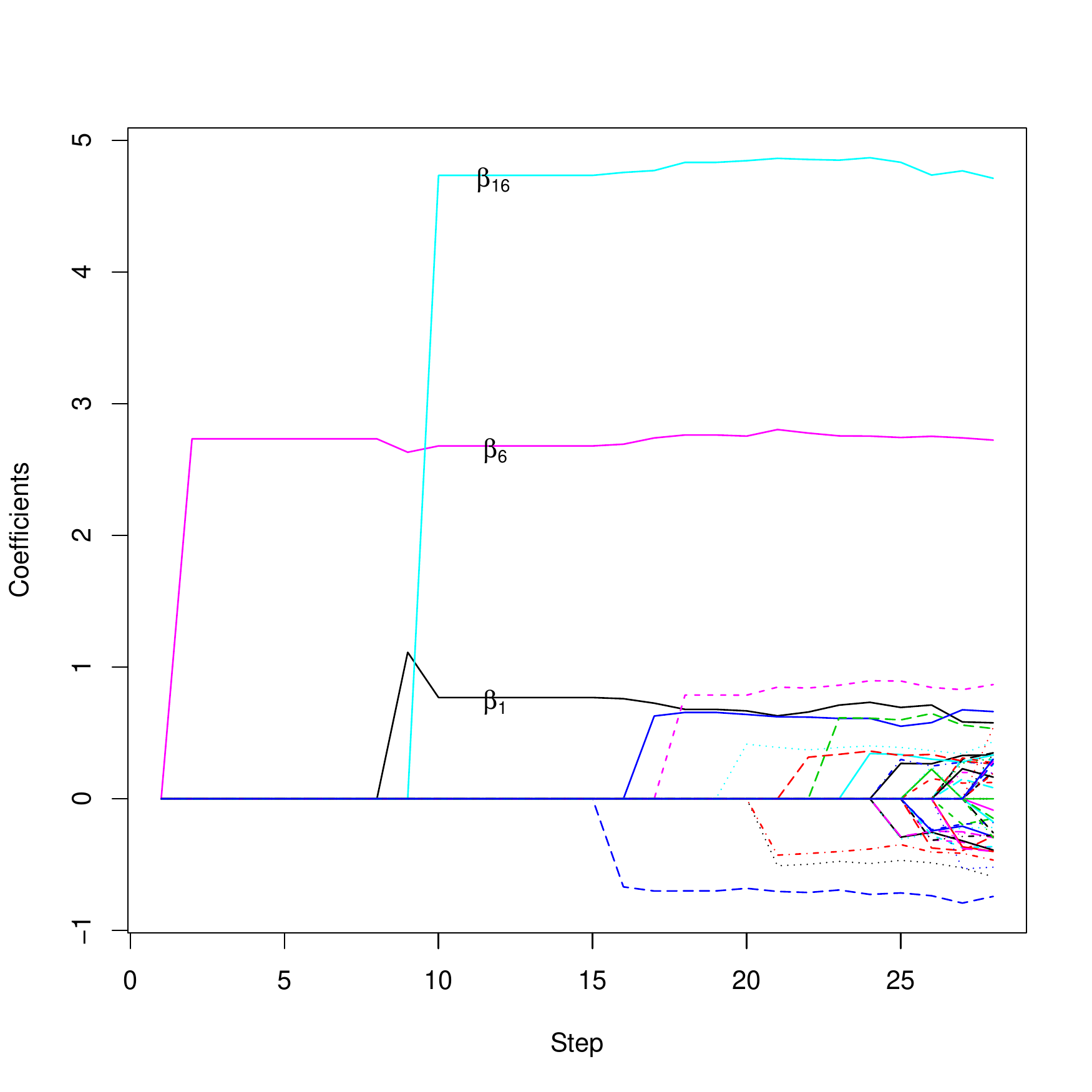}                    %for pdf figure
\includegraphics[scale=0.4]{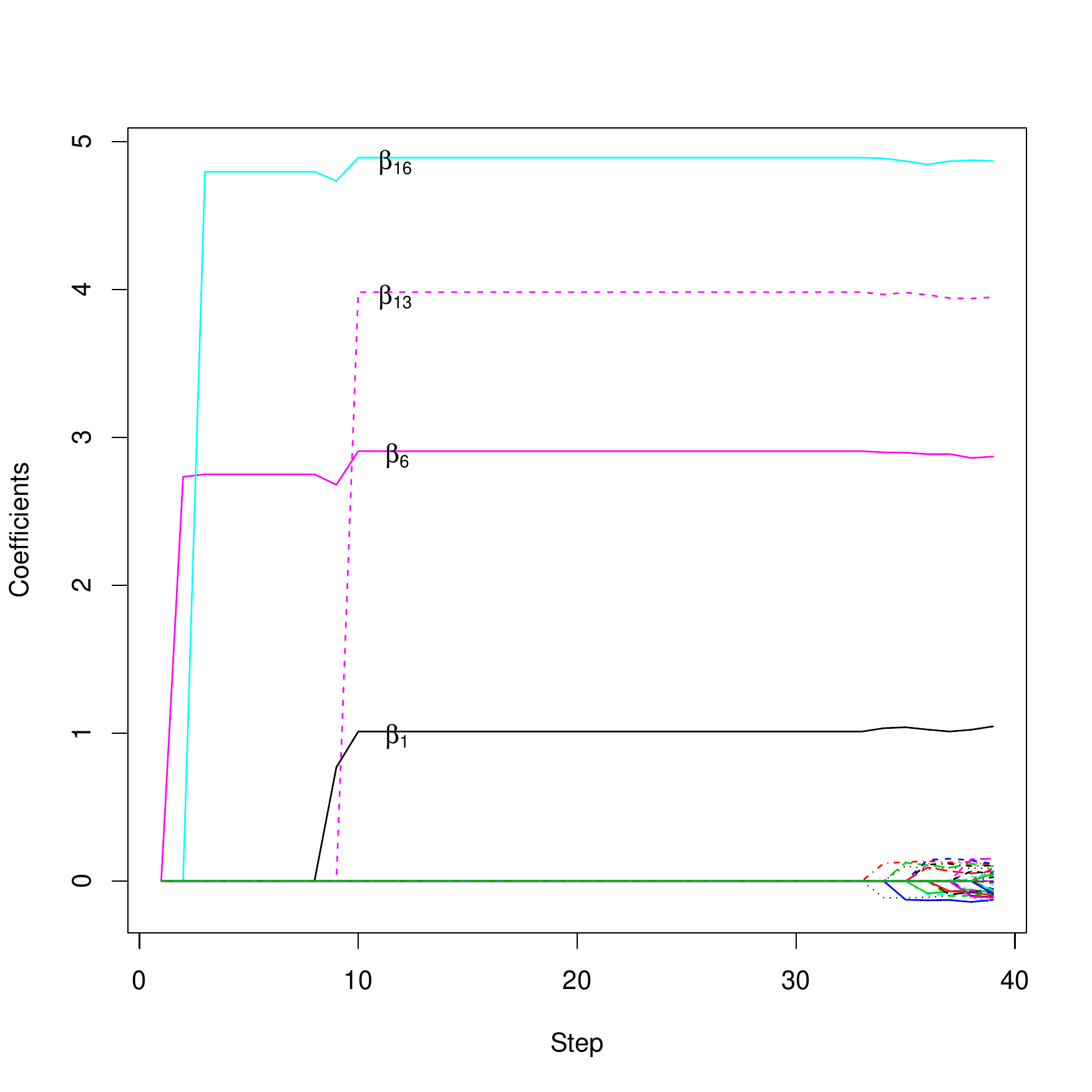}                  %for pdf figure
\end{center}
\vspace{-0.2 in}
\caption{\blue{Two} hierarchy-preserving solution \blue{paths} for a toy example \blue{produced by the RAMP and RAMP-w, respectively}. Left Panel: \blue{strong hierarchy}. Right Panel: \blue{weak hierarchy}.}
\label{fig::path2}
\end{figure}

\red{Figure \ref{fig::path2} illustrates} \blue{two} \red{hierarchy-preserving solution paths obtained by} \blue{the} \red{RAMP} \blue{under} \red{strong and weak heredity constraints}, \blue{respectively}. \red{In this toy example, $n=500, p=100$, and $X_{ij}\stackrel{i.i.d.}{\sim} \mathcal{N}(0,1)$},  \blue{and} \red{$Y=X_1 +3X_6+4X_1X_3+5X_1X_6 +\epsilon$, where $\epsilon\sim \cN(0,1)$.  %There are two important main effects, $X_1$ and $X_6$,  and two important interaction effects, $X_1X_3$ and $X_1X_6$.
Without the marginality principle}, \blue{the interaction term} \red{$X_1X_6$ would be the most significant predictor as it has the highest marginal correlation with the response $Y$. On the other hand, RAMP with the strong heredity selects $X_1$ and $X_6$ before picking up $X_1X_6$ on the solution path. Note that RAMP does not select the interaction $X_1X_3$ until at a very late stage on the solution path due to the strong heredity assumption}.
%($X_3$ is not important as a main effect).
\blue{Under} \red{the weak heredity assumption}, \blue{the RAMP-w} is able to select in sequence $X_6$, $X_1X_6$, $X_1$ and $X_1X_3$. The reason is that after $X_6$ is selected, $X_1X_6$ is \blue{immediately added to} \red{the candidate interaction set and} \blue{then} \red{selected, even before $X_1$ is selected}. %since $X_1X_6$ has a large interaction effect in the model.
\blue{Similarly, the interaction} \red{$X_1X_3$ is picked up by the algorithm after one of its parents $X_1$ is picked up. }

%Figure \ref{fig::path2} illustrates a hierarchy-preserving solution path obtained by RAMP. In this example, $n=400, p=1000$, and $X_{ij}\stackrel{i.i.d.}{\sim} \mathcal{N}(0,1)$,  $Y=2X_1 +3X_3+X_6+5X_1X_3+4X_1X_6 +\epsilon$, where $\epsilon\sim \cN(0,1)$.  There are three important main effects, $X_1$, $X_3$ and $X_6$,  and two important interaction effects, $X_1X_3$ and $X_1X_6$. Without the marginality principle, $X_1X_3$ would be the most significant predictor as it has the highest marginal correlation with the response $Y$. On the other hand, the proposed algorithm always obeys the marginality principle by including $X_1$ and $X_3$ first before selecting $X_1X_3$. In addition, note that the interaction effect $X_1X_3$ is selected before the main effect $X_6$ on the solution path which indicates that the path is different from any two-stage methods.
%
%\begin{figure}[H] %figure 1
%\begin{center}
%\includegraphics[width=6.2in,height=3.8in]{toy-path.pdf}                    %for pdf figure
%%\includegraphics[width=72mm,height=162mm,angle=270]{NewEX11}          %for eps figure
%\end{center}
%\vspace{-0.2 in}
%\caption{Hierarchy-preserving solution path for a toy example.}
%\label{fig::path2}
%\end{figure}

%We have developed a new R package \texttt{RAMP} for the algorithm.

%In S2, strictly speaking, the solution does not necessarily satisfy the strong heredity condition, however, it is an efficient approximation and the strong heredity condition is generally satisfied from our limited numerical experience.

\section{Extension to Generalized Quadratic Regression Models}\label{S4}

\subsection{\red{Generalized Quadratic Regression}}
A standard \red{generalized linear model (GLM)} assumes that the conditional distribution of $\by$ given $\bX$ belongs to the canonical exponential family with density
\begin{eqnarray*}
f_n(\by,\bX,\bbeta)=\prod_{i=1}^{n}f_0(y_i;\theta_i)=\prod_{i=1}^{n}\left\{c(y_i)\exp\left[\frac{y_i\theta_i-b(\theta_i)}{\phi}\right]\right\},
\end{eqnarray*}
where $\phi>0$ is a dispersion parameter, $\bbeta=(\beta_1,...,\beta_p)^{\top}$ are the regression coefficients, and
%$\{f_0(y,\theta):\theta\in\mathbb{R}\}$ is a family of distributions in the regular exponential family with dispersion parameter $\phi>0$
\begin{eqnarray}\label{AAA9}
\btheta=(\theta_1,...,\theta_n)^{\top}=\bX\bbeta.
\end{eqnarray}
The function $b(\btheta)$ is twice continuously differentiable with a positive second-order derivative. In sparse high dimensional modeling, $\bbeta$ is a long vector with a small number of nonzero entries. In the context of QR, the design matrix is $(\bX,\bX^{\circ2})$. A natural generalization of GLM is to modify (\ref{AAA9}) as
\begin{eqnarray}\label{AAA10}
\btheta=(\theta_1,...,\theta_n)^{\top}=\bX\bbeta_{\cM}+\bX^{\circ2}\bbeta_{\cI}.
\end{eqnarray}
\blue{In the literature}, \red{there are} \blue{very} \red{few computational tools available to fit high dimensional GQR models. Next, we illustrate how the aforementioned algorithms can be used} \blue{for} \red{GQR.}
%where $\bbeta_{\cI}$ is a $p(p+1)/2$ vector indexed by $\cI$.

\subsection{Two-stage Regularization Methods}
For high dimensional data, the penalized likelihood method is commonly used to fit GLM. Given the systematic component (\ref{AAA9}), the penalized likelihood estimator is defined as
\begin{eqnarray*} %\label{temp3}
\argmin_{\bbeta}Q_n(\bbeta)=\argmin_{\bbeta}-\ell_n(\bbeta)+\sum_{j=1}^pJ_{\lambda}(|\beta_j|),
\end{eqnarray*}
where $\ell_n(\bbeta)=\log f_n(\by;\bX,\bbeta)=\frac1n\left(\by^{\top}\bX\bbeta-\bone^{\top}\bb(\bX\bbeta)\right)$ is the log-likelihood up to a scalar, $J_{\lambda} (\cdot)$ is a penalty function and $\lambda\geq 0$ is the regularization parameter. %Popular penalties include the LASSO \citep{TibshiraniLASSO:1996}, SCAD \citep{FanLiSCAD:2001}, MCP \citep{ZhangMCP:2010}, among many others. Coordinate optimization method can be used to solve (\ref{temp3}) \citep{park2007l1,friedman2010regularization,FanLvSCAD:2011,Yu.Feng.2013}.

For GQR with systematic component (\ref{AAA10}), we propose the two-stage approach as follows. At Stage 1, only main effects are selected by the penalization method with order-2 terms being left out. Denote the selected main-effect set by $\widehat{\cA}$. At Stage 2, we expand $\widehat\cA$ by adding all the two-way interactions (children) of those main effects (parents) within $\widehat\cA$ and solve
\begin{eqnarray*}%\label{temp4}
\argmin_{\bbeta}Q_n(\bbeta, \widehat\cA)=\argmin_{\bbeta}-\ell_n(\bbeta, \widehat\cA)+\sum_{\alpha\in\widehat{\cA}^{\circ2}} J_{\lambda}(\beta_{\alpha}),
\end{eqnarray*}
where
\begin{align*}
\ell_n(\bbeta, \widehat\cA)=\frac{1}{n}\left[\by^T(\bX_{\widehat{\cA}}\bbeta_{\widehat{\cA}}+\bX_{\widehat{\cA}}^{\circ2}\bbeta_{\widehat{\cA}^{\circ2}})-\bone^{\top}\bb(\bX_{\widehat{\cA}}\bbeta_{\widehat{\cA}}+\bX_{\widehat{\cA}}^{\circ2}\bbeta_{\widehat{\cA}^{\circ2}})\right].
\end{align*}

At Stage 2, we intentionally do not impose penalty on main effects in $\widehat\cA$, so that all the selected main effects at Stage 1 will stay in the final model. This will assure the hierarchical structure of main effects and interactions in the final model.

\subsection{New Path Algorithm for Generalized QR}
The RAMP proposed in Section 3 can be easily extended to fit the GQR. The major difference is to replace the penalized least squares by the penalized likelihood function at each step. The CDA algorithm is used to minimize the penalized likelihood function iteratively.

\bigskip
\noindent
{\bf RAMP Algorithm for GQR}:

\vspace{2mm}
{\it Initialization}: Set $\lambda_{\max}=n^{-1}\max |\bX^{\top}\by|$ and  $\lambda_{\min}=\zeta\lambda_{\max}$ with $1>\zeta>0$. Generate an exponentially decaying sequence $\lambda_{\max}=\lambda_1>\lambda_2>\cdots>\lambda_K=\lambda_{\min}$. Initialize the main effect set $\cM_{0}=\emptyset $ and the interaction effect set $\cI_{0}=\emptyset$.

\vspace{3mm}
{\it Path-building}: Repeat the following steps for $k=1,\cdots,K$. Given $\cM_{k-1}, \cI_{k-1}, \cH_{k-1}$, add the possible interactions among main effects in $\cM_{k-1}$ to the current model. Then with respect to $(\beta_0,\bbeta^{\top},\bbeta_{\cM_{k-1}^{\circ2}}^{\top})^{\top}$, we maximize
%\begin{eqnarray}\label{temp50}
%\ell_n(\bbeta, \cM_{k-1})-\lambda_k\sum_{j\in {\cH}^c_{{k-1}}} |\beta_j|- \lambda_k\|\bbeta_{\cM_{k-1}^{\circ2}}\|_1,
%\end{eqnarray}
\begin{eqnarray*}%\label{temp50}
\ell_n(\bbeta, \cM_{k-1})-\lambda_k\|\bbeta_{\cH^c_{{k-1}}}\|_1- \lambda_k\|\bbeta_{\cM_{k-1}^{\circ2}}\|_1,
\end{eqnarray*}
where
\begin{align*}
\ell_n(\bbeta, \cM_{k-1})=&\frac{1}{n}\left[\by^T(\bX_{{\cM_{k-1}}}\bbeta_{{\cM_{k-1}}}+\bX_{{\cM_{k-1}}}^{\circ2}\bbeta_{\cM_{k-1}^{\circ2}})\right.\\
&\left.-\bone^{\top}\bb(\bX_{{\cM_{k-1}}}\bbeta_{{\cM_{k-1}}}+\bX_{{\cM_{k-1}}}^{\circ2}\bbeta_{\cM_{k-1}^{\circ2}})\right].
\end{align*}
Calculate $\cM_k$, $\cI_k$ and $\cH_k$ according to the solution. Add the corresponding main effects from $\cI_k$ into $\cM_k$ to enforce the heredity constraint, and calculate the MLE based on the current model.

\section{Numerical Studies}\label{S5}

\subsection{Simulation Examples}
\red{We consider data generating processes with varying signal-to-noise ratios, different covariate structures, error distributions, and heredity structures. In particular, Example \ref{ex:linear-p=5000} is a QR model} \blue{under a} \red{$p\gg n$ settings with strong heredity considered in \cite{HaoZhang:forward:2012}. Example \ref{ex:logi-p=2000}} \blue{is a high-dimensional} \red{logistic regression} \blue{model} \red{with interaction effects. Examples \ref{ex::linear-p=100} and \ref{ex:linear-p=200} consider QR models with} \blue{the} \red{weak and strong heredity structures respectively,} \blue{where we consider} \red{a relatively small $p$ to make the comparison possible with} \blue{the} \red{hierarchical lasso \citep{bien2013lasso}.  %Example \ref{ex:linear-p=200} is a lower dimensional version of Example \ref{ex:linear-p=5000} to enable the comparison with hierarchical lasso \citep{bien2013lasso}.
Example \ref{ex:linear-p=5000-t-error} considers a QR model with a heavy tail error distribution to demonstrate the robustness of our} \blue{methods}.

\blue{For comparison, } \red{we consider} \blue{RAMP and} \red{two two-stage methods, i.e.,  two-stage LASSO (2-LASSO) and two-stage SCAD (2-SCAD). We also include existing methods iFORT and iFORM \citep{HaoZhang:forward:2012}}, \blue{the} \red{hierarchical lasso \citep{bien2013lasso}, and the benchmark method ORACLE for which the true sparse model is known.}

\blue{When computing} \red{the solution} \blue{paths} \red{of two-stage methods and RAMP, we choose the tuning parameter by EBIC with $\gamma=1$ \citep{chenchen:2008}. We} \blue{also implemented other parameter tuning criteria} including AIC, BIC, and GIC \citep{fan2013tuning}, and \blue{observed that} \red{the EBIC tends to work the best among most of the simulation settings that we considered. For easy presentation, we report only the results for EBIC.}

Let $\cS=\{j:\beta_j\ne0\}$ and $\cT=\{(j,k):\beta_{j,k}\ne0\}$ with cardinality $s=|\cS|$ and $t=|\cT|$. For each example, we run $M=100$ Monte-Carlo simulations for each method and make a comparison. For the $m$-th simulation, denote the estimated subsets as $\widehat\cS^{(m)}$ and $\widehat\cT^{(m)}$, the estimated coefficient vector as $\widehat\bbeta^{(m)}$, the main effects and interaction effects as $\widehat\beta^{(m)}_j$ and $\widehat\beta^{(m)}_{j,k}$. We evaluate the performance on variable selection and model estimation %and prediction
based on the following criteria.
\begin{itemize}
\item Main effects coverage percentage (main.cov): $M^{-1}\sum_{m=1}^M I(\cS\subset \widehat\cS^{(m)})$.
\item Interaction effects coverage percentage (inter.cov): $M^{-1}\sum_{m=1}^M I(\cT\subset \cTm)$.
\item Main effects exact selection percentage (main.exact): $M^{-1}\sum_{m=1}^M I(\cS= \cSm)$.
\item Interaction effects exact selection percentage (inter.exact): $M^{-1}\sum_{m=1}^M I(\cT= \cTm)$.
\item Model size (size): $M^{-1}\sum_{m=1}^M (|\cSm|+|\cTm|)$.
\item Root mean squared error (RMSE):
$\{M^{-1}\sum_{m=1}^M[\sum_{j=0}^p(\hat\beta_j^{(m)}-\beta_j)^2+\sum_{j=1}^p\sum_{k=j}^p(\hat\beta_{j,k}^{(m)}-\beta_{j,k})^2]\}^{1/2}$.
%\item Out-of-sample $R^2$ ($R^2$):
%the average prediction performance on a test data set of size $n$.
\end{itemize}

%\begin{example}\label{ex::1}
%We generate data from model \eqref{AAA1} with $(n,p,p_0,q_0)=(200,500,4,4)$. $\bX\stackrel{i.i.d.}{\sim} MVN(\bzero, I_p)$. $\bbeta_{\cM}=(3,0,3,0,0,3,0,0,0,3,\bzero_{490})^T$. The important interaction effects are $\cT=\{(1,3), (1,6), (3,10), (6,10)\}$ with coefficient 2.
%\end{example}
%
%\begin{example}\label{ex::2}
%The setup is the same as Example \ref{ex::1} except $\bX\stackrel{i.i.d.}{\sim} MVN(\bzero, \bSigma)$, where
%$\Sigma_{jk}=0.5^{|j-k|}$.
%\end{example}
\begin{example}\label{ex:linear-p=5000}
\normalfont
Set $(n,p,s,t)=(400,5000,10,10)$. Generate the covariates $\{\bx_i\}_{i=1}^n\stackrel{i.i.d.}{\sim} \cN(\bzero, \bSigma)$ with
$\Sigma_{jk}=0.5^{|j-k|}$ and generate the response $\by$ by model \eqref{AAA1}. $\cS=\{1,2,\cdots,10\}$ with the true regression coefficients $\bbeta_{\cS}=(3,3,3,3,3,2,2,2,2,2)^{\top}$. The set of important interaction effects is $\cT=\{(1,2),(1,3),(2,3),(2,5),(3,4),(6,8),(6,10),(7,8),(7,9),(9,10)\}$ with the corresponding coefficients $(2, 2, 2, 2, 2, 1, 1, 1, 1, 1)$.
\end{example}

%\begin{example}\label{ex:linear-p=10000}
%	Set $(n,p,s,t)=(400,10000,10,10)$. The rest setup is same as Example \ref{ex:linear-p=5000}.
%\end{example}

\begin{table}[t]
\caption{Selection and estimation results  for Example \ref{ex:linear-p=5000}. \label{tb::ex3}}
\centering
\begin{tabular}{l|l|cc|cc|cc}
\hline
&&\multicolumn{2}{c|}{main effects} &\multicolumn{2}{c|}{interaction effects} &&\\
&$\sigma$&coverage &exact &coverage & exact &size&RMSE\\
\hline
\multirow{3}{*}{RAMP}
&$2$&1.00&0.96&1.00&0.35&20.98&0.87\\
&$3$&0.99&0.91&0.83&0.17&21.25&1.29\\
&$4$&0.92&0.77&0.47&0.11&20.83&1.96\\
\hline
\multirow{3}{*}{2-LASSO}
&$2$&0.78&0.60&0.78&0.01&24.77&1.56\\
&$3$&0.75&0.56&0.75&0.01&24.64&1.85\\
&$4$&0.72&0.51&0.69&0.01&24.40&2.20\\
\hline
\multirow{3}{*}{2-SCAD}
&$2$&0.70&0.58&0.70&0.53&19.92&1.81\\
&$3$&0.69&0.55&0.62&0.26&20.31&2.06\\
&$4$&0.65&0.52&0.43&0.14&20.56&2.42\\
\hline
\multirow{3}{*}{iFORT}&$2$&0.00 &      0.00 &      0.00 &      0.00 &     14.54 &      6.64  \\
&$3$& 0.00 &      0.00 &      0.00 &      0.00 &     13.74 &      7.02  \\
&$4$&0.00 &      0.00 &      0.00 &      0.00 &     12.72 &      7.52 \\
\hline
\multirow{3}{*}{iFORM}&$2$&1.00 &      0.98 &      0.98 &      0.40 &     20.71 &      0.59\\
&$3$&1.00 &      0.97 &      0.34 &      0.17 &     19.94 &      1.40 \\
&$4$& 0.97 &      0.97 &      0.02 &      0.01 &     18.71 &      2.16 \\
\hline
\multirow{3}{*}{ORACLE}&$2$& 1.00 &      1.00 &      1.00 &      1.00 &     20.00 &      0.55 \\
&$3$&1.00 &      1.00 &      1.00 &      1.00 &     20.00 &      0.83 \\
&$4$& 1.00 &      1.00 &      1.00 &      1.00 &     20.00 &      1.11 \\
\hline
\end{tabular}
\end{table}

To have different signal-to-noise ratio situations, we consider $\sigma \in \{2, 3, 4\}$. The results are summarized in Table \ref{tb::ex3}. With regard to model selection, the proposed RAMP has a high coverage percentage in selecting both main effects and interaction effects. The 2-LASSO tends to miss some important main effects while picking up some noise variables, ending up with the largest model size on average. On the other hand, the 2-SCAD has a high exact selection percentage with a low coverage percentage.
Compared to RAMP, the iFORM tends to have a lower coverage on interaction effects. The iFORT is the worst in terms both variable selection and model estimation. With regard to parameter estimation, RAMP has the smallest root mean square error (RMSE) when $\sigma=3$ and 4.
%and the largest out-of-sample $R^2$.

%\begin{table}[t]
%\caption{Selection and estimation results for Example \ref{ex:linear-p=10000}. \label{tb::ex4}}
%\centering
%\begin{tabular}{l|l|cc|cc|cc}
%\hline
%&&\multicolumn{2}{c|}{main effects} &\multicolumn{2}{c|}{interaction effects } &&\\
%&$\sigma$&coverage &exact&coverage & exact &size&RMSE\\
%\hline
%\multirow{3}{*}{RAMP}
%&$2$&1.00&0.94&1.00&0.36&21.04&0.89\\
%&$3$&0.99&0.91&0.78&0.17&21.13&1.34\\
%&$4$&0.73&0.54&0.36&0.04&20.21&2.41\\
%\hline
%\multirow{3}{*}{2-LASSO}&$2$&0.79&0.60&0.79&0.00&24.29&1.55\\
%&$3$&0.75&0.65&0.75&0.00&24.02&1.91\\
%&$4$&0.70&0.63&0.67&0.00&23.76&2.26\\
%\hline
%\multirow{3}{*}{2-SCAD}&$2$&0.71&0.56&0.71&0.51&20.05&1.84\\
%&$3$&0.66&0.56&0.60&0.25&20.47&2.15\\
%&$4$&0.63&0.57&0.42&0.12&20.60&2.55\\
%\hline
%\multirow{3}{*}{iFORT}&$2$&0.00 &      0.00 &      0.00 &      0.00 &     14.11 &      6.72 \\
%&$3$& 0.00 &      0.00 &      0.00 &      0.00 &     13.69 &      7.04 \\
%&$4$&0.00 &      0.00 &      0.00 &      0.00 &     12.40 &      7.60 \\
%\hline
%\multirow{3}{*}{iFORM}&$2$& 1.00 &      0.97 &      0.96 &      0.31 &     20.74 &      0.62 \\
%&$3$&1.00 &      0.99 &      0.29 &      0.10 &     19.55 &      1.53 \\
%&$4$&   0.94 &      0.94 &      0.00 &      0.00 &     18.34 &      2.29 \\
%\hline
%\multirow{3}{*}{ORACLE}&$2$&  1.00 &      1.00 &      1.00 &      1.00 &     20.00 &      0.56 \\
%&$3$& 1.00 &      1.00 &      1.00 &      1.00 &     20.00 &      0.85 \\
%&$4$&1.00 &      1.00 &      1.00 &      1.00 &     20.00 &     1.10 \\
%\hline
%\end{tabular}
%\end{table}

\begin{example}\label{ex:logi-p=2000}
\normalfont
We consider a logistic regression model with $$\log\frac{P(Y=1|X)}{P(Y=0|X)}=\beta_1X_1+3X_6+3X_{10}+3X_1X_6+3X_6X_{10},$$
where $(n, p, s, t)=(400, 2000, 3, 2)$ and
$\bX\stackrel{i.i.d.}{\sim} \cN(\bzero, \bI_p)$. For different signal-to-noise ratios, we vary the coefficient $\beta_1\in \{1, 2, 3\}$.
\end{example}

The results are summarized in Table \ref{tb::ex5}, which lead to the following observations. When the signal is strong ($\beta_1=2, 3$),  RAMP, 2-LASSO and 2-SCAD perform similarly in selecting main effects; while RAMP and 2-SCAD is  much better in selecting interactions than 2-LASSO. When the signal is weak ($\beta_1=1$), 2-LASSO and 2-SCAD fail to identify the correct main effects most of time, which in turn leads to low coverage of important interaction effects. On the other hand, RAMP performs reasonably well in terms of selecting both main effects and interaction effects. With regard to RMSE,  RAMP outperforms 2-LASSO and 2-SCAD in all scenarios. Note that the iFORT and iFORM are omitted in this example, as they do not handle binary responses.

\begin{table}[t]
\caption{Selection and estimation results for Example \ref{ex:logi-p=2000}.  \label{tb::ex5}}
\centering
\begin{tabular}{l|l|cc|cc|cc}
\hline
&&\multicolumn{2}{c|}{main effects} &\multicolumn{2}{c|}{interaction effects } &&\\
&$\beta_{1}$&coverage&exact&coverage&exact&size&RMSE\\
\hline
\multirow{3}{*}{RAMP}
&$1$&0.92&0.78&0.92&0.91&4.98&1.80\\
&$2$&1.00&0.93&1.00&1.00&5.08&1.16\\
&$3$&1.00&0.92&0.99&0.99&5.13&1.36\\
\hline
\multirow{3}{*}{2-LASSO}&
$1$&0.45&0.41&0.45&0.14&4.05&3.97\\
&$2$&1.00&0.93&1.00&0.29&6.58&1.41\\
&$3$&1.00&0.80&1.00&0.42&6.31&1.66\\\hline
\multirow{3}{*}{2-SCAD}&
$1$&0.49&0.43&0.49&0.49&3.58&3.76\\
&$2$&1.00&0.81&1.00&0.94&5.28&1.03\\
&$3$&1.00&0.74&1.00&0.86&5.52&1.22\\
\hline
\multirow{3}{*}{ORACLE}
&$1$&  1.00 &      1.00 &      1.00 &      1.00 &     5.00 &      0.84 \\
&$2$& 1.00 &      1.00 &      1.00 &      1.00 &     5.00 &      0.78 \\
&$3$&1.00 &      1.00 &      1.00 &      1.00 &     5.00 &     0.83 \\
\hline
\end{tabular}
\end{table}

In the next two examples, we compare RAMP and hierNet algorithms for both strong and weak hierarchy scenarios.

\red{
\begin{example}\label{ex::linear-p=100}
\normalfont
Set $(n,p,s,t)=(400,100,10,10)$. Generate the covariates $\{\bx_i\}_{i=1}^n\stackrel{i.i.d.}{\sim} \cN(\bzero, \bSigma)$ with
$\Sigma_{jk}=0.5^{|j-k|}$ and generate the response $\by$ by model \eqref{AAA1}. $\cS=\{1,2,\cdots,10\}$ with the true regression coefficients $\bbeta_{\cS}=(3,3,3,3,3,2,2,2,2,2)^{\top}$. The set of important interaction effects is $\cT=\{(1,2),(1,13),(2,3),(2,15),(3,4),(6,10),(6,18),(7,9),(7,18),(10,19)\}$ with the corresponding coefficients $(2, 2, 2, 2, 2, 1, 1, 1, 1, 1)$.
\end{example}
 In this example, the strong heredity does not hold while the weak heredity is satisfied. Note that we take $p$ to be relatively small due to the heavy computational cost of hierNet \citep{bien2013lasso}. Here, we compare RAMP and RAMP-w (RAMP with} \blue{the} \red{weak heredity constraint) with hierNet-s and hierNet-w}, \blue{and} \red{ the results} \blue{are} \red{summarized in Table \ref{tb::linear-p=100}. As expected, when applying RAMP with strong heredity (RAMP), it always misses some important interaction effects. However, the RAMP with weak heredity (RAMP-w) successfully recovers the important interaction effects with} \blue{a high} \red{proportion, especially when the error variance is small. Comparing with} \blue{the} \red{hierNet,  the RAMP-w} \blue{in general selects} \red{ a much smaller model with a smaller RMSE. In particular, the computation time of hierNet is much longer than RAMP for both the strong and weak versions.
}

\begin{table}[t]
\caption{Selection and estimation results as well as average computing time (in seconds) per replicate for Example \ref{ex::linear-p=100}. \label{tb::linear-p=100}}
\centering
\begin{tabular}{l|l|cc|cc|ccc}
\hline
&&\multicolumn{2}{c|}{main effects} &\multicolumn{2}{c|}{interaction effects } &&\\
&$\sigma$&coverage&exact  &coverage& exact &size&RMSE&Time\\
\hline
\multirow{3}{*}{RAMP}
&$2$&1.00&0.71&0.00&0.00&19.45&3.54&37.49\\
&$3$&1.00&0.83&0.00&0.00&16.86&3.71&34.74\\
&$4$&0.98&0.89&0.00&0.00&15.28&3.87&34.88\\
\hline
\multirow{3}{*}{RAMP-w}
&$2$&1.00&1.00&0.99&0.25&21.33&0.79&47.02\\
&$3$&1.00&0.99&0.63&0.12&21.16&1.31&46.51\\
&$4$&1.00&0.98&0.16&0.00&20.07&1.98&46.10\\
\hline
\multirow{3}{*}{hierNet-s}
&$2$&1.00&0.00&1.00&0.00&133.45&5.69&3143.30\\
&$3$&1.00&0.00&0.96&0.00&119.62&5.33&3232.62\\
&$4$&1.00&0.00&0.74&0.00&95.06&5.01&3507.85\\
\hline
\multirow{3}{*}{hierNet-w}
&$2$&1.00&0.00&1.00&0.00&126.83&6.60&295.88\\
&$3$&1.00&0.01&0.98&0.00&96.59&6.17&346.83\\
&$4$&1.00&0.04&0.75&0.00&65.31&5.73&444.99\\
\hline
\end{tabular}
\end{table}

\red{\begin{example}\label{ex:linear-p=200}
\normalfont
Set $(n,p,s,t)=(400,200,10,10)$. The rest setup is same as Example \ref{ex:linear-p=5000}.
\end{example}
\begin{table}[t]
\caption{Selection and estimation results as well as average computing time (in seconds) per replicate for Example \ref{ex:linear-p=200}. \label{tb::ex7}}
\centering
\begin{tabular}{l|l|cc|cc|ccc}
\hline
&&\multicolumn{2}{c|}{main effects} &\multicolumn{2}{c|}{interaction effects } &&\\
&$\sigma$&coverage&exact&coverage& exact&size&RMSE&Time\\
\hline
\multirow{3}{*}{RAMP}
&$2$&1.00&1.00&1.00&0.35&20.97&0.86&34.58\\
&$3$&1.00&0.98&0.93&0.23&21.31&1.18&32.95\\
&$4$&0.97&0.92&0.64&0.10&21.35&1.72&32.28\\
\hline
\multirow{3}{*}{RAMP-w}
&$2$&1.00&1.00&0.99&0.25&21.25&0.87&56.01\\
&$3$&1.00&1.00&0.78&0.18&21.14&1.25&54.58\\
&$4$&1.00&1.00&0.30&0.06&20.02&1.92&53.71\\
\hline
\multirow{3}{*}{hierNet-s}
&$2$&1.00&0.00&1.00&0.00&120.99&5.53&15847.28\\
&$3$&1.00&0.00&0.99&0.00&115.69&5.15&16552.18\\
&$4$&1.00&0.00&0.92&0.00&90.55&4.79&16864.49\\
\hline
\multirow{3}{*}{hierNet-w}
&$2$&1.00&0.01&1.00&0.00&97.62&5.79&1467.46\\
&$3$&1.00&0.02&0.98&0.00&61.04&5.41&1798.27\\
&$4$&1.00&0.01&0.90&0.00&53.31&5.24&2156.99\\\hline
\end{tabular}
\end{table}
In this example, we} \blue{consider} \red{the case where the strong heredity} \blue{holds} \red{and compare RAMP and RAMP-w with hierNet-s and hierNet-w. From Table \ref{tb::ex7}, it is clear that RAMP outperforms RAMP-w in terms of} \blue{both} \red{the coverage percentage} \blue{and} \red{the exact selection percentage for interaction effect. This is not surprising as the RAMP-w searches for additional interaction effects compared with RAMP. In addition, the RMSE of RAMP is the smallest among the four methods throughout all noise levels. Both hierNet-s and hierNet-w have very good coverage percentage but with almost zero exact selection percentage for both main effects and interaction effects. As a result, they} \blue{select} \red{a large number of noise variables in the final model. Note that the computation time for hierNet-s is over 4 hours for a single replicate. As a result, we omit the comparison with hierNet for the other} \blue{higher dimensional} \red{examples}.
\begin{example}\label{ex:linear-p=5000-t-error}
\normalfont
\red{We use the same setting as in Example \ref{ex:linear-p=5000} except} \blue{for} \red{the error distribution}, \blue{which is changed} \red{to a $t$ distribution with} \blue{degrees} \red{of freedom 3}.
\end{example}
\red{This example is designed to examine the robustness of proposed} \blue{methods under} heavy tail \red{error distributions. For brevity, we report only the performance of the vanilla RAMP with strong heredity enforced. It is clear from Table \ref{tb::ex6} that under the heavy tail error distribution, RAMP has a similar performance} \blue{as in} \red{Example 1.}
\begin{table}[t]
\caption{Selection and estimation results for Example \ref{ex:linear-p=5000-t-error}. \label{tb::ex6}}
\centering
\begin{tabular}{l|l|cc|cc|cc}
\hline
&&\multicolumn{2}{c|}{main effects} &\multicolumn{2}{c|}{interaction effects} &&\\
&$\sigma$&coverage &exact  &coverage& exact&size&RMSE\\
\hline
\multirow{3}{*}{RAMP}
&$2$&1.00&0.94&0.98&0.29&21.59&1.02\\
&$3$&0.97&0.92&0.84&0.18&21.37&1.52\\
&$4$&0.90&0.76&0.49&0.08&21.00&2.33\\
\hline
\end{tabular}
\end{table}
\subsection{Real Data Example: Supermarket Data}
\red{We consider the supermarket dataset analyzed in \cite{wang2009forward} and \cite{HaoZhang:forward:2012}. The data} \blue{set} \red{contains the daily sale information of a major supermarket located in northern China, with $n=464$ and $p=6,398$}.  \blue{The total number of interaction effects is about $2.0\times 10^7$}.
\red{The response $Y$ is the number of customers on a particular day with}
\blue{the} \red{predictor $X$ measuring sale volumes of a selection of products. The supermarket manager would like to find out which products are most informative in predicting the response, which} \blue{would be} \red{useful to design promotions around those products.}

\red{Here, we randomly split the data into} \blue{a} \red{training set ($n_1=400$) and} \blue{a} \red{test set ($n_2=64$) to evaluate the prediction performance of different methods. We} \blue{also} \red{compare the performance of RAMP with the regular LASSO without taking interaction effects into account. Because of the issue} \blue{of tuning parameter selection}, \red{we report the results using different tuning methods including AIC, BIC, EBIC \citep{chenchen:2008}, and GIC \citep{fan2013tuning}} \blue{for} \red{both RAMP and the LASSO. }

\red{For each random split, we calculate the number of selected variables, the number of selected interaction effects, and the out-of-sample $R^2$ on the test set. The average performance over 100 random splits is presented in Table \ref{tb::supermarket}. When we use BIC, EBIC and GIC, RAMP selects a model with higher out-of-sample $R^2$} \blue{values} \red{than the LASSO. When using more stringent tuning parameter criteria like the EBIC and GIC}, \blue{it is observed that the RAMP performs significantly} \red{better than the LASSO. For example}, \blue{when} \red{GIC is used, RAMP selects 30 variables on average with around 3 of them being interaction effects, and has an average out-of-sample $R^2$ value of 90.08, which is much higher than the corresponding} \blue{LASSO results}. \red{It is clear that by using RAMP with the inclusion of possible interaction effects, we can} \blue{obtain} \red{a more interpretable model with a reasonably good prediction performance. Moreover, from Table 8 in \cite{HaoZhang:forward:2012}, the out-of-sample $R^2$ values with the associated standard error for iFORT and iFORM are 88.91 (0.17) and 88.66 (0.18), respectively, both of which are outperformed by RAMP with any tuning parameter selection method. }

\begin{table}
\caption{Mean selection and prediction results on the supermarket data \blue{set} over 100 random splits. The standard errors are in parentheses. \label{tb::supermarket}}
\begin{tabular}{c|ccc|ccc}\hline
&\multicolumn{3}{c|}{RAMP}&\multicolumn{3}{c}{LASSO}\\
\hline&size&size.inter&$R^2$&size&size.inter&$R^2$\\\hline
AIC&229.12(1.68)&94.53(1.06)&90.48(0.23)&264.28(0.91)&0.00(0.00)&92.04(0.18)\\
BIC&101.17(3.25)&34.36(1.65)&91.18(0.20)&63.47(0.77)&0.00(0.00)&90.76(0.20)\\
EBIC&29.27(1.01)&3.07(0.29)&89.67(0.31)&15.62(0.46)&0.00(0.00)&72.09(0.53)\\
GIC&30.71(0.92)&3.20(0.30)&90.08(0.28)&19.19(0.74)&0.00(0.00)&75.05(0.58)\\
\hline
\end{tabular}
	
\end{table}

\section{Discussion}\label{S6}
We study regularization methods for interaction selection subject to the marginality principle for QR and GQR models. \red{One main advantage of these algorithms is their computational efficiency and feasibility for high and ultra-high dimensional data. In particular, a key feature of RAMP is that it can select main and interaction effects simultaneously while still keeping the hierarchy structure. The strategy of RAMP can be used to extend other algorithms, e.g., LARS, to build the entire solution path when fitting the regularized QR models. %This paper focuses only on the LASSO-type estimator due to its simple form of the penalty function. The proposed algorithms can be implemented to incorporate other penalty functions such as SCAD and MCP.
All algorithms considered in this paper utilize the hierarchy} \blue{structures. Such structures are} \red{natural for quadratic models \citep{Nelder:1977,HaoZhang:note:2012}. Nevertheless, in certain applications, some main effects may not be strong enough to be selected first without incorporating the interaction effects}. \blue{Other} \red{approaches \citep{ZhaoETAL:2009, YuanRoshanZou:2009, ChoiLiZhu:2010,bien2013lasso} can be applied in this scenario, as these methods keep the hierarchy in different ways. However, a drawback is that most of these algorithms are relatively slow when $p$ is large.  Recently, there} \blue{have been studies} on interaction selection \red{which do not rely on the strong or weak hierarchy.  Based on the idea of sure independence screening \citep{fan2008sure,  fan2011nonparametric, cheng2014nonparametric},  \cite{jiang2014variable} proposed Sliced Inverse Regression for Interaction Detection (SIRI) for screening interaction variables;  \cite{fan2016interaction} introduced a new approach called interaction pursuit for interaction identification using screening and variable selection. It would be interesting to  incorporate} \red{these} screening based methods into our path algorithm to handle general \blue{scenarios.}
%It is a promising research direction to develop new scalable interaction selection techniques that are less dependent on the hierarchy.

\red{We demonstrate theoretical properties of the two-stage LASSO method for QR. As a referee pointed out, selection consistency results on the LASSO often rely on the irrepresentable condition}, \blue{which} \red{is not realistic in applications. In order to extend current results, it is desirable to investigate a broad range of penalty functions for GQR, e.g., under frameworks similar to \cite{FanLvSCAD:2011} and \cite{fan2013asymptotic}. }

%\red{The success of the proposed regularization methods depend highly on the strong/weak heredity. If neither heredity is satisfied, the proposed methods are not guaranteed to work well. There has been several recent work on this case.  Based on the idea of sure independence screening \citep{fan2008sure,  fan2011nonparametric, cheng2014nonparametric},  \cite{jiang2014variable} proposed Sliced Inverse Regression for Interaction
%Detection (SIRI) for screening important interaction variables, and recently  \cite{fan2016interaction} introduced a new approach, called interaction pursuit, for interaction identification using screening and variable selection. It would be interesting to  incorporate the screening based methods into our path algorithm to handle the general scenario. }

%\red{Theory: Possible extensions, 1. other penalties to get rid of irrepresentable condition; 2. GQR; 3. non-Gaussian design. Methodology: 1. hierarchical method or not (heredity conditions) 2. weak hierarchy 3. Screening consistency is enough 4. Parameter selection.}

An R package \texttt{RAMP} has been developed and is available from the CRAN website.

%\section*{Acknowledgement}
%\red{The authors are partially supported by NSF Grants DMS-1309507 (Hao and Zhang), DMS-1308566 and DMS-1554804 (Feng). The authors are grateful to the editors, associate editor, and three referees for their helpful comments and suggestions.}

\section{Appendix}
The main results are shown in Appendix A, and a related lemma is put in Appendix B.
\subsection{Appendix A}
%We use most notations in \cite{wainwright2009sharp} with some exceptions. Here $s=|\cS|$ is the number of true important main effects in the data generating process, which is denoted by $k$ in \cite{wainwright2009sharp}. Vectors and matrices are expressed in boldface. The $n$-vector $\bomega$ is a random Gaussian noise in \cite{wainwright2009sharp}; but in our paper it is the imaginary noise at Stage 1, i.e., the sum of the Gaussian noise $\bv$ and the interaction effects $(\bu_1^{\top}\bbeta_{\cI},...,\bu_n^{\top}\bbeta_{\cI})^{\top}$. So it is not independent of the design matrix $\bX$. However, under Gaussian design or any design with certain symmetry, $\bomega$ is uncorrelated with the design matrix, as argued in \cite{HaoZhang:note:2012}.

{\bf Proof of Theorem 1.}  We will apply the primal-dual witness (PDW) method and use $(W1)$, $(W2)$, etc. to denote the formula $(1)$, $(2)$,... in \cite{wainwright2009sharp}. Recall in our paper, the $n$-vector $\bomega$ is the imaginary noise at Stage 1, which is the sum of the Gaussian noise $\bv$ and the interaction effects $(\bu_1^{\top}\bbeta_{\cI},...,\bu_n^{\top}\bbeta_{\cI})^{\top}$, and hence it is not independent of the design matrix $\bX$. %Since $\bomega$ here is not independent of design matrix, we need to fix any part involving $\bomega$.

\vspace{2mm}
\noindent
{\it Part I: Verifying strict dual feasibility}.

The goal is to show that, with overwhelming probability, under condition (\ref{AAA6}), inequality $|Z_j|<1$ holds for each $j\in\cS^c$, where $Z_j$ is defined in (W10).
For every $j\in\cS^c$, conditional on $\bX_{\cS}$, (W37) gives a decomposition $Z_j=A_j+B_j$ where
\begin{eqnarray*}
A_j&=&\bE_j^{\top}\left\{\bX_{\cS}(\bX_{\cS}^{\top}\bX_{\cS})^{-1}\check{\bz}_{\cS}+\Pi_{\bX_{\cS}^{\perp}}\left(\frac{\bomega}{\lambda_n n}\right)\right\}\\
B_j&=&\Sigma_{j\cS}(\Sigma_{\cS\cS})^{-1}\check{\bz}_{\cS},
\end{eqnarray*}
where $\bE_j^{\top}=\bX_j^{\top}-\Sigma_{j\cS}(\Sigma_{\cS\cS})^{-1}\bX_{\cS}^{\top}\in\mathbb{R}^n$ with $E_{ij}\sim\cN(0,[\Sigma_{\cS^c|\cS}]_{jj})$.

Condition (C1) implies
\begin{eqnarray*}
\max_{j\in\cS^c}|B_j|\leq 1-\gamma.
\end{eqnarray*}

Conditioned on $\bX_{\cS}$ and $\bomega$, $A_j$ is Gaussian with mean zero and variance $\Var(A_j)\leq\rho_u(\Sigma_{\cS^c|\cS})M_n$ where
\[M_n=\frac1n\check\bz^{\top}_{\cS}\left(\frac{\bX_{\cS}^{\top}\bX_{\cS}}{n}\right)^{-1}\check\bz_{\cS}+\left\|\Pi_{\bX_{\cS}^{\perp}}\left(\frac{\bomega}{\lambda_n n}\right)\right\|_2^2.\]

The following lemma, proved in appendix B, generalizes Lemma 4 in \cite{wainwright2009sharp}.

\begin{lemma}
For any $\epsilon\in(0,\frac12)$, define the event $\overline\cT(\epsilon)=\{M_n>\overline{M}_n(\epsilon)\}$, where
\begin{eqnarray*}
\overline{M}_n(\epsilon)=\left(1+\max\left\{\epsilon,\frac{8}{C_{\min}}\sqrt{\frac{s}{n}}\right\}\right)\left(\frac{s}{C_{\min}n}+\frac{2(\sigma^2+\tau^2)}{\lambda_n^2n}\right).
\end{eqnarray*}
Then $\bP(\overline\cT(\epsilon))\leq C_1\exp(-C_2\min\{\sqrt{n}\epsilon^2,s\})$ for some $C_1,$ $C_2>0$.
\end{lemma}

By Lemma 1,
\begin{eqnarray}
\nonumber \bP\left(\max_{j\in\cS^c}|Z_j|\geq1\right)&\leq&\bP\left(\max_{j\in\cS^c}|A_j|\geq\gamma\right)\\
&\leq&\bP\left(\max_{j\in\cS^c}|A_j|\geq\gamma\mid\overline\cT^c(\epsilon))\right)+C_1\exp(-C_2\min\{\sqrt{n}\epsilon^2,s\})\label{AAA11}.
\end{eqnarray}
Note that the goal is to show the probability in (\ref{AAA11}) is exponentially decayed. Conditional on $\overline\cT^c(\epsilon)$, $\Var(A_j)\leq\rho_u(\Sigma_{\cS^c|\cS})\overline{M}_n(\epsilon)$, so
\begin{eqnarray*}
\bP\left(\max_{j\in\cS^c}|A_j|\geq\gamma\mid\overline\cT^c(\epsilon))\right)\leq 2(p-s)\exp\left(-\frac{\gamma^2}{2\rho_u(\Sigma_{\cS^c|\cS})\overline{M}_n(\epsilon)}\right).
\end{eqnarray*}

The assumptions of \red{Theorem} 1 imply $\frac{s}{n}=o(1)$ and $\frac{1}{\lambda_n^2n}=o(1)$, so $\overline{M}_n(\epsilon)=o(1)$. Therefore, it suffices to show that the decaying rate of the exponential term dominates $p-s$. It is easy to check that (\ref{AAA6}) can guarantee that $\max_{j\in\cS^c}|Z_j|<1$ holds with probability at least $1-c_1\exp(-c_2\min\{s,\log(p-s),n^{\frac12}\})$.

Now we show the sufficiency of the alternative condition (\ref{AAA8}). In particular,we show (\ref{AAA5}) and (\ref{AAA8}) imply (\ref{AAA6}), which is equivalent to
\[\frac{n}{1+\delta}>2s\log(p-s)\frac{\rho_u(\Sigma_{\cS^c|\cS})}{C_{\min}\gamma^2}(1+\frac{2(\sigma^2+\tau^2)C_{\min}}{\lambda^2_ns}).\]
Plugging in (\ref{AAA5}), we have
\begin{eqnarray}
\nonumber \frac{n}{1+\delta}&>&2s\log(p-s)\frac{\rho_u(\Sigma_{\cS^c|\cS})}{C_{\min}\gamma^2}+2s\log(p-s)\frac{\rho_u(\Sigma_{\cS^c|\cS})}{C_{\min}\gamma^2}\frac{2(\sigma^2+\tau^2)C_{\min}}{\lambda^2_ns}\\
&=&2s\log(p-s)\frac{\rho_u(\Sigma_{\cS^c|\cS})}{C_{\min}\gamma^2}+\frac{n}{\phi_p}\frac{\log(p-s)}{\log p}.\label{AAA12}
\end{eqnarray}
Following the same argument after (W40) in \cite{wainwright2009sharp}, (\ref{AAA12}) is implied by (\ref{AAA8}) for $\phi_p\geq 2$.

%The first part of the proposition follows by PDW method.
\medskip
\noindent
{\it Part II: Sign consistency}.

In order to show sign consistency, we need to show that (W13) holds. That is
\begin{eqnarray}\label{AAA13}
\sign(\beta_j+\Delta_j)=\sign(\beta_j), \quad \text{for all } j\in\cS,
\end{eqnarray}
where
\begin{eqnarray*}
\Delta_j=\be_j^{\top}\left(\frac{\bX_{\cS}^{\top}\bX_{\cS}}{n}\right)^{-1}\left[\frac1n\bX_{\cS}^{\top}\bomega-\lambda_n\sign(\bbeta_{\cS})\right].
\end{eqnarray*}

From definition, we have
\begin{eqnarray*}
\max_{j\in\cS}|\Delta_j|\leq F_1+F_2\leq F_1+(F_{2,1}+F_{2,2}),
\end{eqnarray*}
where
\begin{eqnarray*}
F_1&=&\lambda_n\left\|\left(\frac{\bX_{\cS}^{\top}\bX_{\cS}}{n}\right)^{-1}\sign(\bbeta_{\cS})\right\|_{\infty}\\
F_2&=&\left\|\left(\frac{\bX_{\cS}^{\top}\bX_{\cS}}{n}\right)^{-1}\frac1n\bX_{\cS}^{\top}\bomega\right\|_{\infty}\\
F_{2,1}&=&\left\|\left(\frac{\bX_{\cS}^{\top}\bX_{\cS}}{n}\right)^{-1}\frac1n\bX_{\cS}^{\top}\bv\right\|_{\infty}\\
F_{2,2}&=&\left\|\left(\frac{\bX_{\cS}^{\top}\bX_{\cS}}{n}\right)^{-1}\frac1n\bX_{\cS}^{\top}\by_{\cI}\right\|_{\infty}
\end{eqnarray*}

(W41) and a correction version of (W42) give upper bounds of tail probability of $F_1$ and $F_{2,1}$, respectively. That is
\begin{eqnarray}\label{AAA14}
\bP\left(F_1>c_3\lambda_n\left\|\Sigma_{\cS\cS}^{-\frac12}\right\|_{\infty}^2\right)\leq 4\exp(-c_2\min\{s,\log(p-s)\}),
\end{eqnarray}
\begin{eqnarray}\label{AAA15}
\bP\left(F_{2,1}\geq 20\sqrt{\frac{\sigma^2s}{C_{\min} n}}\right)\leq 4\exp(-c_1 s).
\end{eqnarray}
Now we work on the addition term $F_{2,2}$. By (W60),
\[\bP\left(\left\|\left(\frac1n \bX_{\cS}^{\top}\bX_{\cS}\right)^{-1} \right\|_2\geq \frac{9}{C_{\min}}\right) \leq 2\exp(-n/2).\]
\[\left\|\frac1n\bX_{\cS}^{\top}\by_{\cI}\right\|_2\leq \|\bbeta_{\cI}\|_2\max_{j\in\cS;(k,\ell)\in\cT}\left\{\left|\frac1n\bX_j^{\top}(\bX_k\star\bX_{\ell})\right|\right\}.\]
$\frac1n\bX_j^{\top}(\bX_k\star\bX_{\ell})$ is a sample third moment, so by Lemma \red{B.5} in \cite{HaoZhang:forward:2012},
\[\bP\left(\left|\frac1n\bX_j^{\top}(\bX_k\star\bX_{\ell})\right|>\epsilon\right)\leq c_4\exp(-c_5n^{\frac23}\epsilon^2).\]
Therefore,
we have
\[\bP\left(\left\|\frac1n\bX_{\cS}^{\top}\by_{\cI}\right\|_2\geq  \|\bbeta_{\cI}\|_2\epsilon\right)\leq s^3 c_4\exp(-c_5n^{\frac23}\epsilon^2).\]
Overall,
\[\bP\left(F_{2,2}\geq \frac{9}{C_{\min}}\|\bbeta_{\cI}\|_2\epsilon \right)\leq  s^3 c_6\exp(-c_7n^{\frac23}\epsilon^2).\]
Setting $\epsilon=\frac{s^{\frac12}}{n^{\frac13}}$,
we have
\begin{eqnarray}\label{AAA16}
\bP\left(F_{2,2}\geq \frac{9\|\bbeta_{\cI}\|_2\sqrt{s }}{C_{\min}n^{\frac13}}  \right)\leq  c_8\exp(-c_9s).
\end{eqnarray}
Combining (\ref{AAA14}), (\ref{AAA15}) and (\ref{AAA16}), we have that with probability greater than $1-c'_1\exp(-c'_2\\\min\{s,\log(p-s)\})$,
\[\max_{j\in\cS}|\Delta_j|\leq c_3\lambda_n\left\|\Sigma_{\cS\cS}^{-\frac12}\right\|_{\infty}^2+20\sqrt{\frac{\sigma^2s }{C_{\min} n}}+\frac{9\|\bbeta_{\cI}\|_2\sqrt{s }}{C_{\min}n^{\frac13}}=g(\lambda_n).\]
Therefore (\ref{AAA13}) holds when $\beta_{\min}>g(\lambda_n)$.
$\Box$

\subsection{Appendix B}
{\bf Proof of Lemma 1.}
The first summand of $M_n$ can be controlled exactly the same way as in \cite{wainwright2009sharp}, i.e.,
\[\frac1n\check\bz^{\top}_{\cS}\left(\frac{\bX_{\cS}^{\top}\bX_{\cS}}{n}\right)^{-1}\check\bz_{\cS}\leq\left(1+\frac{8}{C_{\min}}\sqrt{\frac{s}{n}}\right)\frac{s}{nC_{\min}}\]
with probability at least $1-2\exp(-s/2)$.

Turning to the second summand, we observe that $\Pi_{\bX_{\cS}^{\perp}}$ is an orthogonal projection matrix and $\bomega=\bv+\by_{\cI}$, so
\[\left\|\Pi_{\bX_{\cS}^{\perp}}\left(\frac{\bomega}{\lambda_n n}\right)\right\|_2^2\leq\frac{\|\bomega\|_2^2}{\lambda^2_nn^2}\leq\frac{2}{\lambda_n^2n}\frac{\|\bv\|_2^2+\|\by_{\cI}\|_2^2}{n}.\]
%$\|\bomega\|_2^2\leq 2(\|\bv\|_2^2+\|\by_{\cI}\|_2^2)$.
Note that $\|\bv\|_2^2/\sigma^2\sim \chi^2_n$, by (W54a),
\begin{eqnarray}\label{AAA17}
\bP\left(\frac{\|\bv\|_2^2}{n}\leq(1+\epsilon) {\sigma^2} \right)\leq \exp\left(-\frac{3n\epsilon^2}{16}\right).
\end{eqnarray}
Moreover,
\[\|\by_{\cI}\|^2_2-n\tau^2=\sum_{i=1}^n(\bu_i^{\top}\bbeta_{\cI})^2-\tau^2,\]
is a sum of mean zero independent random variables. Define $\bB=(B_{jk})$ is the coefficient matrix with $B_{jk}=\beta_{j,k}/2$, ($j\ne k$) and $B_{jj}=\beta_{j,j}$.%, and $\bB$ is sparse and supported in $\bB_{\cS\cS}$. %by the strong heredity condition.

For each $i$, we can write %remove the heredity condition
%\[\bu_i^{\top}\bbeta_{\cI}=\bx_i^{\top}\bB\bx_i-\E(\bx_i^{\top}\bB\bx_i)=(\bx_i)_{\cS}^{\top}\bB_{\cS\cS}(\bx_i)_{\cS}-\E\left((\bx_i)_{\cS}^{\top}\bB_{\cS\cS}(\bx_i)_{\cS}\right)
%=\be_i^{\top}\bA\be_i-\tr(\bA),\]
%where $\be_i\sim\cN(\bzero,\bI_{s\times s})$, $\bA=(\Sigma_{\cS\cS})^{\frac12}\bB_{\cS\cS}(\Sigma_{\cS\cS})^{\frac12}$.
\[\bu_i^{\top}\bbeta_{\cI}=\bx_i^{\top}\bB\bx_i-\E(\bx_i^{\top}\bB\bx_i)=\be_i^{\top}\bA\be_i-\tr(\bA),\]
where $\be_i\sim\cN(\bzero,\bI)$, $\bA=(\Sigma)^{\frac12}\bB(\Sigma)^{\frac12}$.

The moment generating function $M(t)$ of the quadratic form $\be_i^{\top}\bA\be_i$ is
\begin{eqnarray}\label{AAA18}
M(t)=\E e^{t\be_i^{\top}\bA\be_i}=\det(\bI-2t\bA)^{-\frac12}=\prod_{j=1}^s(1-2t\lambda_j)^{-\frac12},
\end{eqnarray}
where $\{\lambda_j\}_{j=1}^s$ are eigenvalues of $\bA$ with ascending order. From (\ref{AAA18}), we have
\[\E(\be_i^{\top}\bA\be_i)=\tr(\bA),\quad \Var(\be_i^{\top}\bA\be_i)=2\tr(\bA^2)=\tau^2,\]
and
\[\Var\left( (\be_i^{\top}\bA\be_i-\tr(\bA))^2\right)=48\tr(\bA^4)+8\tr^2(\bA^2).\]
%Define $\sqrt{W_i}=\frac{\be_i^{\top}\bA\be_i-\tr(\bA)}{\tau}$, then $\E(\sqrt{W_i})=0$, $\Var(\sqrt{W_i})=\E(W_i)=1$, $\Var(W_i)=12 \frac{\tr(\bA^4)}{\tr^2(\bA^2)}+2\leq 14$. Moreover,
Define $W_i=\frac{(\be_i^{\top}\bA\be_i-\tr(\bA))^2}{\tau^2}$, then $\E(W_i)=1$, $\Var(W_i)=12 \frac{\tr(\bA^4)}{\tr^2(\bA^2)}+2\leq 14$. Moreover,
\begin{eqnarray*}
\E e^{t|W_i|^{\frac12}}&=&\E e^{t\frac{|\be_i^{\top}\bA\be_i-\tr(\bA)|}{\tau}}\\
                       &\leq&\E e^{t\frac{\be_i^{\top}\bA\be_i-\tr(\bA)}{\tau}}+E e^{-t\frac{\be_i^{\top}\bA\be_i-\tr(\bA)}{\tau}}\\
                       &=& e^{-t\frac{\tr(\bA)}{\tau}}M(\frac{t}{\tau})+e^{t\frac{\tr(\bA)}{\tau}}M(\frac{-t}{\tau})\\
                       &=&\left(\prod_{j=1}^s \frac{e^{-\sqrt{2}ta_j}}{1-\sqrt{2}ta_j}\right)^{\frac12}+\left(\prod_{j=1}^s \frac{e^{\sqrt{2}ta_j}}{1+\sqrt{2}ta_j}\right)^{\frac12}
\end{eqnarray*}
where $a_j=\lambda_j/\sqrt{\sum_{j=1}^s\lambda_j^2}$, so $\sum_{j=1}^sa_j^2=1$. It is easy to see $\frac{e^{-x}}{1-x}\leq 1+x^2$ for $x\in[-\frac12,\frac12]$.
For $0\leq t\leq \frac{\sqrt{2}}{4}$, $|\sqrt{2}ta_j|\leq\frac12$, so both summand in the last formula can be controlled by
\[\left(\prod_{j=1}^s(1+2t^2a_j^2)\right)^{\frac12}\leq\left(\prod_{j=1}^s(1+ a_j^2/4)\right)^{\frac12}\leq\left(\prod_{j=1}^se^{a_j^2/4}\right)^{\frac12}=e^{\frac12\sum_{j=1}^s\frac{a_j^2}{4}}=e^{\frac18}.\]

Therefore, $\E e^{t|W_i|^{\frac12}}\leq 2e^{\frac18}$ for $0\leq t\leq \frac{\sqrt{2}}{4}$. And $\E e^{t|W_i-1|^{\frac12}}\leq \E e^{t(|W_i|+1)^{\frac12}} \leq \E e^{t+t|W_i|^{\frac12}}\leq 2e^{\frac{\sqrt{2}}{4}+\frac18}$.

By \red{Lemma B.4} in \cite{HaoZhang:forward:2012},
\[\bP\left(\left|\sum_{i=1}^n(W_i-1)\right|>n\epsilon \right)\leq c_1\exp(-c_2 n^{\frac12}\epsilon^2),\]
for some positive constants $c_1$, $c_2$. That is
\[\bP \left(\left|\|\by_{\cI}\|^2_2-n\tau^2\right|\geq \tau^2 n\epsilon\right)\leq c_1\exp(-c_2 n^{\frac12}\epsilon^2),\]
which implies
\begin{eqnarray}\label{AAA19}
\bP\left(\frac{\|\by_{\cI}\|_2^2}{n}\leq(1+\epsilon) {\tau^2} \right)\leq c_1\exp\left(-c_2 n^{\frac12}\epsilon^2\right).
\end{eqnarray}
(\ref{AAA17}) and (\ref{AAA19}) imply
\[\bP\left( \left\|\Pi_{\bX_{\cS}^{\perp}}\left(\frac{\bomega}{\lambda_n n}\right)\right\|_2^2\geq(1+\epsilon)\frac{2(\sigma^2+\tau^2)}{\lambda^2_nn}\right)\leq c_3\exp\left(-c_4 n^{\frac12}\epsilon^2\right).\]
And the conclusion of Lemma 1 follows. $\Box$

%\bibliographystyle{biometrika}
%\bibliography{interaction}

\section*{Supplementary of ``Model Selection for High Dimensional Quadratic Regression via Regularization''}
 
\noindent{\bf Supplementary A: Theorem 2}

In this supplementary to our paper \cite{hao2014model}, we show a generalized version of Theorem 1 without Gaussian assumption. Similar as in \cite{hao2014model}, constants $C_1$, $C_2$,... and $c_1$, $c_2$,... are locally defined and may take different values in different sections. We start with a brief review of definition of a subgaussian random variable and its properties.

A random variable $X$ is called $b$-subgaussian if for some $b>0$, $E(e^{tX})\leq e^{b^2t^2/2}$ for all $t\in\mathbb{R}$. The set of all subgaussian random variables is closed under linear operation by the following proposition.

\begin{proposition}\label{PP1}
Let $X_i$ be $b_i$-subgaussian for $i=1,...,n$. Then $a_1X_1+...+a_nX_n$ is $B$-subgaussian with $B=\sum_{i=1}^n|a_i|b_i$. Moreover, if $X_1$,...,$X_n$ are independent, $a_1X_1+...+a_nX_n$ is $B$-subgaussian with $B=\left(\sum_{i=1}^n a_i^2b_i^2\right)^{\frac12}$.
\end{proposition}

Moreover, the tail probability of a subgaussian variable can be well controlled.
\begin{proposition}\label{PP2}
If $X$ is $b$-subgaussian, then $\bP(|X|>t)\leq 2e^{-\frac{t^2}{2b^2}}$ for all $t>0$. Moreover, there exists a positive constant, say $a=1/6b^2$, such that $Ee^{aX^2}\leq 2$.
\end{proposition}
These well-known results can be found, e.g., in \cite{rivasplata2012subgaussian}.

%To generalize Theorem 1, we consider a subgaussian design. That is,

\noindent{\bf Condition (SG)} $\{\bx_i\}_{i=1}^n$ are IID random vectors from an elliptical distribution with marginal $b$-subgaussian distribution. Moreover, $\{\v_i\}_{i=1}^n$ are IID with $b$-subgaussian distribution.

We still use $\Sigma$ and $\Sigma_{\cA\cB}$ denote the covariance matrix of $\bx_i$ and its submatrix corresponding to index sets $\cA$ and $\cB$. $\bB=(B_{jk})$ is the coefficient matrix for interaction effects with $B_{jk}=\beta_{j,k}/2$, ($j\ne k$) and $B_{jj}=\beta_{j,j}$. $\Lambda_{\min}(\bA)$ and $\Lambda_{\max}(\bA)$ denote the smallest and largest eigenvalues of a matrix $\bA$. We need the following technical conditions:
\begin{description}
   \item[(C1)] (Irrepresentable Condition) $\|\Sigma_{\cS^c\cS}(\Sigma_{\cS\cS})^{-1}\|_{\infty}\leq 1-\gamma,$ $\gamma\in (0,1]$.
   \item[(C2)] (Eigenvalue Condition) $\Lambda_{\min}(\Sigma_{\cS\cS})\geq C_{\min}>0$.
   \item[(C3)] (Dimensionality and Sparsity) $s\log p=o(n)$ and $s(\log s)^{\frac12}=o(n^{\frac13})$.
   \item[(C4)] (Coefficient Matrix) $\bB$ is sparse and supported in a submatrix $\bB_{\cS\cS}$. $\Lambda_{\max}(\bB^2)=\Lambda_{\max}(\bB^2_{\cS\cS})\leq C^2_{\bB}$ for a positive constant $C_{\bB}$.
\end{description}

Condition (C3) is employed to replace (6) in Theorem 1. Similar conditions are standard in the literature. Condition (C4) on $\bB$ is used to control the overall interaction effect, which is treated as noise in stage one.  $\Lambda_{\max}(\bB)$ can be bounded, e.g., by $\|\bbeta_{\cI}\|_1$. %$\sum_{j,k}|\beta_{j,k}|$.

\begin{theorem} \label{thm2}
Suppose that conditions (SG), (C1)-(C4) hold. For $\lambda_n\gg \tau\left(\log p/n\right)^{\frac12}$, with probability tending to 1, the LASSO has a unique solution $\hat\bbeta_L$ with support contained within $\cS$.
Moreover, if $\beta_{\min}=\min_{j\in\cS}|\beta_j|>2(s^{-\frac12}+\|\bbeta_{\cI}\|_2/s+\lambda_ns^{\frac12})/C_{\min}$, then $\sign(\hat\bbeta_L)=\sign(\bbeta_{\cM})$.
\end{theorem}

Note that $\|\bbeta_{\cI}\|_2=\tr(B^2)\leq sC^2_{\bB}$, so $\|\bbeta_{\cI}\|_2/s\leq C_{\bB}s^{-\frac12}$.

\noindent{\bf Supplementary B: Proof of Theorem 2}

\noindent Recall that we use $(W1)$, $(W2)$,... to denote the formula $(1)$, $(2)$,... in \cite{wainwright2009sharp}. The $n$-vector $\bomega$ is the imaginary noise at Stage 1, which is the sum of the subgaussian noise $\bv$ and the interaction effects $(\bu_1^{\top}\bbeta_{\cI},...,\bu_n^{\top}\bbeta_{\cI})^{\top}$.

\noindent
{\it Part I: Verifying strict dual feasibility}.

We show that  inequality $|Z_j|<1$ holds for each $j\in\cS^c$, with overwhelming probability, where $Z_j$ is defined in (W10).
For every $j\in\cS^c$, conditional on $\bX_{\cS}$, (W37) gives a decomposition $Z_j=A_j+B_j$ where
\begin{eqnarray*}
A_j&=&\bE_j^{\top}\left\{\bX_{\cS}(\bX_{\cS}^{\top}\bX_{\cS})^{-1}\check{\bz}_{\cS}+\Pi_{\bX_{\cS}^{\perp}}\left(\frac{\bomega}{\lambda_n n}\right)\right\}\\
B_j&=&\Sigma_{j\cS}(\Sigma_{\cS\cS})^{-1}\check{\bz}_{\cS},
\end{eqnarray*}
where $\bE_j^{\top}=\bX_j^{\top}-\Sigma_{j\cS}(\Sigma_{\cS\cS})^{-1}\bX_{\cS}^{\top}\in\mathbb{R}^n$ with entries $E_{ij}$ that is $2b$-subgaussian by Proposition \ref{PP1} and condition (C1).

Condition (C1) implies
\begin{eqnarray*}
\max_{j\in\cS^c}|B_j|\leq 1-\gamma.
\end{eqnarray*}

Conditional on $\bX_{\cS}$ and $\bomega$, $A_j$ is $2bM_n^{\frac12}$-subgaussian, where  %Implicitly, elliptical distribution is need here as we are dealing with CONDITIONAL.... Moreover, we need another fact that a conditional subgaussian variable has smaller aubgaussian norm than the unconditional one.
\[M_n=\frac1n\check\bz^{\top}_{\cS}\left(\frac{\bX_{\cS}^{\top}\bX_{\cS}}{n}\right)^{-1}\check\bz_{\cS}+\left\|\Pi_{\bX_{\cS}^{\perp}}\left(\frac{\bomega}{\lambda_n n}\right)\right\|_2^2.\]

We need the following lemma that is proved in Supplementary C.%, generalizes Lemma 4 in \cite{wainwright2009sharp}.

\begin{lemma}\label{LL2}
For any $\epsilon\in(0,\frac12)$, define the event $\overline\cT(\epsilon)=\{M_n>\overline{M}_n(\epsilon)\}$, where
\begin{eqnarray*}
%\overline{M}_n(\epsilon)=\left(1+\max\left\{\epsilon,\frac{8}{C_{\min}}\sqrt{\frac{s}{n}}\right\}\right)\left(\frac{s}{C_{\min}n}+\frac{2(\sigma^2+\tau^2)}{\lambda_n^2n}\right).
\overline{M}_n(\epsilon)=\frac{2s}{C_{\min}n}+\frac{4(\sigma^2+\tau^2)}{\lambda_n^2n}.
\end{eqnarray*}
Then $\bP(\overline\cT(\epsilon))\leq C_1s^2\exp(-C_2n^{\frac12}\epsilon^2)$ for some $C_1,$ $C_2>0$.
\end{lemma}

By Lemma \ref{LL2},
\begin{eqnarray}
\nonumber \bP\left(\max_{j\in\cS^c}|Z_j|\geq1\right)&\leq&\bP\left(\max_{j\in\cS^c}|A_j|\geq\gamma\right)\\
&\leq&\bP\left(\max_{j\in\cS^c}|A_j|\geq\gamma\mid\overline\cT^c(\epsilon))\right)+C_1s^2\exp(-C_2n^{\frac12}\epsilon^2)\label{BBB5}.
\end{eqnarray}
%Note that the goal is to show the probability in (\ref{BBB5}) converges to 0.
Conditional on $\overline\cT^c(\epsilon)$, $A_j$ is $2b\overline M_n^{\frac12}(\epsilon)$-subgaussian, so by Proposition \ref{PP2}
\begin{eqnarray*}
\bP\left(\max_{j\in\cS^c}|A_j|\geq\gamma\mid\overline\cT^c(\epsilon))\right)\leq 2(p-s)\exp\left(-\frac{\gamma^2}{8b^2\overline{M}_n(\epsilon)}\right),
\end{eqnarray*}
where the right hand side goes to 0 by condition (C3). Therefore, $\max_{j\in\cS^c}|Z_j|<1$ holds with probability tending to 1.

\medskip
\noindent
{\it Part II: Sign consistency}.

In order to show sign consistency, by Lemma 3 in \cite{wainwright2009sharp} it is sufficient to show
\begin{eqnarray}\label{BBB7}
\sign(\beta_j+\Delta_j)=\sign(\beta_j), \quad \text{for all } j\in\cS,
\end{eqnarray}
where
\begin{eqnarray*}
\Delta_j=\be_j^{\top}\left(\frac{\bX_{\cS}^{\top}\bX_{\cS}}{n}\right)^{-1}\left[\frac1n\bX_{\cS}^{\top}\bomega-\lambda_n\sign(\bbeta_{\cS})\right].
\end{eqnarray*}

It is straightforward that
\begin{eqnarray*}
\max_{j\in\cS}|\Delta_j|&\leq&  \left\|\left(\frac{\bX_{\cS}^{\top}\bX_{\cS}}{n}\right)^{-1}\right\|_2\left\|\frac1n\bX_{\cS}^{\top}\bomega-\lambda_n\sign(\bbeta_{\cS})\right\|_2 \\
                                        &\leq&  \left\|\left(\frac{\bX_{\cS}^{\top}\bX_{\cS}}{n}\right)^{-1}\right\|_2\left(\left\|\frac1n\bX_{\cS}^{\top}\bv\right\|_2+\left\|\frac1n\bX_{\cS}^{\top}\by_{\cI}\right\|_2+\left\|\lambda_n\sign(\bbeta_{\cS})\right\|_2 \right).
\end{eqnarray*}
By Lemma \ref{LL3},
\[\left\|\left(\frac{\bX_{\cS}^{\top}\bX_{\cS}}{n}\right)^{-1}\right\|_2<2/C_{\min},\] with probability at least $1-s^2C_3\exp(-C_4n/s^2)$.
Moreover, \[\left\|\lambda_n\sign(\bbeta_{\cS})\right\|_2 \leq \lambda_ns^{\frac12}.\]

\[\left\|\frac1n\bX_{\cS}^{\top}\by_{\cI}\right\|_2\leq \|\bbeta_{\cI}\|_2\max_{j,k,\ell\in\cS}\left\{\left|\frac1n\bX_j^{\top}(\bX_k\star\bX_{\ell})\right|\right\},\]
where $\frac1n\bX_j^{\top}(\bX_k\star\bX_{\ell})$ is a sample third moment. By Remark B.2 and Lemma B.5 in \cite{HaoZhang:forward:2012},
\[\bP\left(\left|\frac1n\bX_j^{\top}(\bX_k\star\bX_{\ell})\right|>\epsilon\right)\leq c_1\exp(-c_2n^{\frac23}\epsilon^2).\]
Because $|\cS|=s$,
we have
\[\bP\left(\left\|\frac1n\bX_{\cS}^{\top}\by_{\cI}\right\|_2\geq  \|\bbeta_{\cI}\|_2\epsilon\right)\leq s^3 c_1\exp(-c_2n^{\frac23}\epsilon^2).\]
which, with $\epsilon=1/s$ leads to
\[\bP\left(\left\|\frac1n\bX_{\cS}^{\top}\by_{\cI}\right\|_2\geq  \|\bbeta_{\cI}\|_2s^{-1}\right)\leq s^3 c_1\exp(-c_2n^{\frac23}/s^2).\]

Similarly,
\[\bP\left(\left\|\frac1n\bX_{\cS}^{\top}\bv\right\|_2>s^{\frac12}\epsilon\right)< s c_3\exp(-c_4n\epsilon^2),\]
which, with $\epsilon=1/s$ leads to
\[\bP\left(\left\|\frac1n\bX_{\cS}^{\top}\bv\right\|_2>s^{-\frac12} \right)< s c_3\exp(-c_4n/s^2).\]

Overall, with probability greater than $1-c_5s^3\exp(-c_6 n^{\frac23}/s^2)$,
\[\max_{j\in\cS}|\Delta_j|\leq 2\left(s^{-\frac12}+\|\bbeta_{\cI}\|_2s^{-1}+\lambda_ns^{\frac12}\right)/C_{\min}=g(\lambda_n).\]
Therefore (\ref{BBB7}) holds when $\beta_{\min}>g(\lambda_n)$.
$\Box$

\noindent{\bf Supplementary C: Proof of Lemma \ref{LL2}.}

\noindent The first summand of $M_n$ can be bounded as
%\[\frac1n\check\bz^{\top}_{\cS}\left(\frac{\bX_{\cS}^{\top}\bX_{\cS}}{n}\right)^{-1}\check\bz_{\cS}\leq\left(1+\frac{8}{C_{\min}}\sqrt{\frac{s}{n}}\right)\frac{s}{nC_{\min}}\]
\[\frac1n\check\bz^{\top}_{\cS}\left(\frac{\bX_{\cS}^{\top}\bX_{\cS}}{n}\right)^{-1}\check\bz_{\cS}\leq \frac{2s}{nC_{\min}}\]
with probability at least $1-s^2C_3\exp(-C_4n/s^2)$, where $C_3$, $C_4$ are positive constants. It directly follows the fact $\|\check\bz_{\cS}\|^2_2\leq s$ and Lemma \ref{LL3} in Supplementary D, which says the largest eigenvalue of $\left(\frac{\bX_{\cS}^{\top}\bX_{\cS}}{n}\right)^{-1}$ can be controlled by $2/C_{\min}$.

For the second summand, because $\Pi_{\bX_{\cS}^{\perp}}$ is an orthogonal projection matrix and $\bomega=\bv+\by_{\cI}$, we have
\[\left\|\Pi_{\bX_{\cS}^{\perp}}\left(\frac{\bomega}{\lambda_n n}\right)\right\|_2^2\leq\frac{\|\bomega\|_2^2}{\lambda^2_nn^2}\leq\frac{2}{\lambda_n^2n}\frac{\|\bv\|_2^2+\|\by_{\cI}\|_2^2}{n}.\]
%$\|\bomega\|_2^2\leq 2(\|\bv\|_2^2+\|\by_{\cI}\|_2^2)$.

%Note that $\|\bv\|_2^2/\sigma^2\sim \chi^2_n$, by (W54a),
As $\{\bv_i\}_{i=1}^n$ are IID subgaussian, by Proposition \ref{PP2}, and Lemma B.4 in \cite{HaoZhang:forward:2012}, we have
\begin{eqnarray}\label{BBB11}
\bP\left(\frac{\|\bv\|_2^2}{n}\leq(1+\epsilon) {\sigma^2} \right)\leq c_1\exp\left(-c_2n\epsilon^2\right).
\end{eqnarray}
On the other hand,
\[\|\by_{\cI}\|^2_2-n\tau^2=\sum_{i=1}^n(\bu_i^{\top}\bbeta_{\cI})^2-\tau^2,\]
is a sum of mean zero independent random variables.

Define $W_i=\frac{(\bu_i^{\top}\bbeta_{\cI})^2}{\tau^2}-1$, then $\E(W_i)=0$. By condition (C4),
\[\bu_i^{\top}\bbeta_{\cI}=\bx_i^{\top}\bB\bx_i-\E(\bx_i^{\top}\bB\bx_i)=(\bx_i)_{\cS}^{\top}\bB_{\cS\cS}(\bx_i)_{\cS}-\E\left((\bx_i)_{\cS}^{\top}\bB_{\cS\cS}(\bx_i)_{\cS}\right).\]
So $W_i$ is a degree 4 polynomial of subgaussian variables dominated by $[C_{\bB}(\bx_i)_{\cS}^{\top}(\bx_i)_{\cS}]^2$, which is, up to the constant $C_{\bB}^2$, a summation of at most $s^2$ degree 4 monomials of subgaussian variables. The tail probability of each of these monomials can be bounded as in Lemma B.5 in \cite{HaoZhang:forward:2012}. Therefore, we have

\[\bP\left(\left|\sum_{i=1}^nW_i\right|>n\epsilon \right)\leq c_3s^2\exp(-c_4 n^{\frac12}\epsilon^2),\]
for some positive constants $c_3$, $c_4$. That is
\[\bP \left(\left|\|\by_{\cI}\|^2_2-n\tau^2\right|\geq \tau^2 n\epsilon\right)\leq c_3s^2\exp(-c_4 n^{\frac12}\epsilon^2),\]
which implies
\begin{eqnarray}\label{BBB13}
\bP\left(\frac{\|\by_{\cI}\|_2^2}{n}\leq(1+\epsilon) {\tau^2} \right)\leq c_3s^2\exp\left(-c_4 n^{\frac12}\epsilon^2\right).
\end{eqnarray}
(\ref{BBB11}) and (\ref{BBB13}) imply
\[\bP\left( \left\|\Pi_{\bX_{\cS}^{\perp}}\left(\frac{\bomega}{\lambda_n n}\right)\right\|_2^2\geq(1+\epsilon)\frac{2(\sigma^2+\tau^2)}{\lambda^2_nn}\right)\leq c_5s^2\exp\left(-c_6 n^{\frac12}\epsilon^2\right),\]
for some positive constants $c_5$, $c_6$. With $\epsilon=1$, the conclusion of Lemma \ref{LL2} follows. $\Box$

\noindent{\bf Supplementary D: Lemma \ref{LL3} and its proof.}

\begin{lemma}\label{LL3}
Under conditions (SG) and (C3), we have \[\bP\left(\Lambda_{\min}\left(\frac{\bX_{\cS}^{\top}\bX_{\cS}}{n}\right)>C_{\min}/2\right)>1-s^2C_3\exp(-C_4n/s^2)\to 1,\] where $C_{\min}=\Lambda_{\min}(\Sigma_{\cS\cS})$, $C_3>0$, $C_4>0$.
\end{lemma}

{\bf Proof.} We need bound
\begin{eqnarray}\label{BBB14}
\bP\left(\sup_{\|\bbv\|_2=1}|\bbv^{\top}(\Sigma_{\cS\cS}- \bX_{\cS}^{\top}\bX_{\cS}/n)\bbv|>\epsilon \right).
\end{eqnarray}
For easy presentation, we assume that the $s$-vector $\bbv$ is indexed by $\cS$. Then
\begin{eqnarray*}
& & |\bbv^{\top}(\Sigma_{\cS\cS}- \bX_{\cS}^{\top}\bX_{\cS}/n)\bbv|\\
&\leq&\sum_{j,k\in\cS}|v_jv_k||\Sigma_{jk}-\bX_j^{\top}\bX_k/n|\\
&\leq&\|\bbv\|_1^2\max_{j,k\in\cS}|\Sigma_{jk}-\bX_j^{\top}\bX_k/n| \\
&\leq& s\max_{j,k\in\cS}|\Sigma_{jk}-\bX_j^{\top}\bX_k/n|
\end{eqnarray*}

So (\ref{BBB14}) is bounded from above by
\begin{eqnarray}\label{BBB15}
\bP\left(\max_{j,k\in\cS}|\Sigma_{jk}-\bX_j^{\top}\bX_k/n|>\epsilon/s \right)
\end{eqnarray}
Following Remark B.2 and Lemma B.5 in \cite{HaoZhang:forward:2012}, it is easy to derive
\[ \bP\left(|\Sigma_{jk}-\bX_j^{\top}\bX_k/n|>\epsilon \right)< C_3\exp(-C_5n\epsilon^2),\]
for constants $C_3>0$, $C_5>0$ under subgaussian assumption. Therefore, (\ref{BBB15}) is further bounded by
$s^2C_3\exp(-C_5n\epsilon^2/s^2)$. Take $\epsilon=\min\{C_{\min}/2,1/2\}$, we have
\[\bP\left(\Lambda_{\min}\left(\frac{\bX_{\cS}^{\top}\bX_{\cS}}{n}\right)>C_{\min}/2\right)>1-s^2C_3\exp(-C_4n/s^2)\to1,\]
by condition (C3), where $C_4=C_5(\min\{C_{\min}/2,1/2\})^2$.

\bibliographystyle{biometrika}
\bibliography{interaction}
\end{document}